# What the Solar System Can Teach Us About Rocky Exoplanets

**Paul K. Byrne[1,2] · Claire Marie Guimond[3,4] · Peter A. Cawood[5] · Michael J. Way[6,7,8] · Doris Breuer[9] · Tilman Spohn[9] · João C. Duarte[10] · Diogo L. Lourenço[4] · Francesca Miozzi[4,11,12] · Maëlis Arnould[13] · Nicolas Coltice[14] · Stephanie L. Olson[15]**

**Abstract**
The number of rocky extrasolar planets being discovered continues to increase, but so too does the apparent diversity of such worlds. As we work to understand the likely bulk properties and thermal, geological, and climatological attributes we might expect of rocky exoplanets, we can look to the Solar System for guidance. Here, we review the interior, surface, and atmospheric characteristics of modern Earth, and discuss how our homeworld has changed through Solar System history. We then visit in turn Venus, Mars, Mercury and Earth's Moon, and Io, noting how these terrestrial planets are alike, how they are different, and how they have evolved through time. Finally, we consider some examples of the types of rocky worlds known or suspected to exist without direct Solar System counterparts—but we argue that even then, for all the variety we might expect, they are still variations on a common theme. Perhaps the most important lesson the Solar System can teach us is that terrestrial bodies change through time, sometimes dramatically, and that rocky worlds of similar size and mass can have vastly different planetary outcomes.

## 1 Introduction

### 1.1 What Art Thou, Rocky Exoplanet?

The first potentially rocky world identified outside the Solar System, the ∼six-Earth-mass ($M_\oplus$) CoRoT-7 b, was not discovered until 2009 (Léger et al. 2009; Queloz et al. 2009), 17 years after the first confirmed exoplanet detection of any kind. Yet our tally of rocky exoplanets is growing, even as we face limits in detector technology and fundamental astrophysical noise limits that dwarf planetary signals (Bouchy et al. 2001; Vanderburg et al. 2016; Klein et al. 2024). We now know that worlds the size of Earth (and, by extension, Venus) are common in our cosmic neighbourhood. This finding, based on three decades of searching for exoplanets, represents how distinguishing our planetary system from others has become an empirical science.

Our planet census is not complete, however, because exoplanet detection quickly becomes more difficult for ever smaller bodies with increasingly tenuous astrophysical signals. And so, to estimate the occurrence rates of planets as a function of their radius, mass,

Extended author information available on the last page of the article

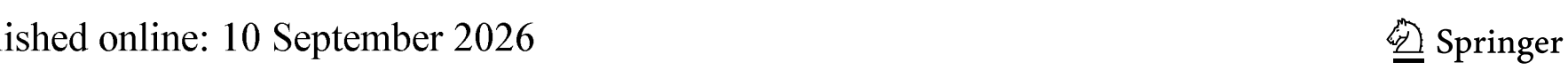

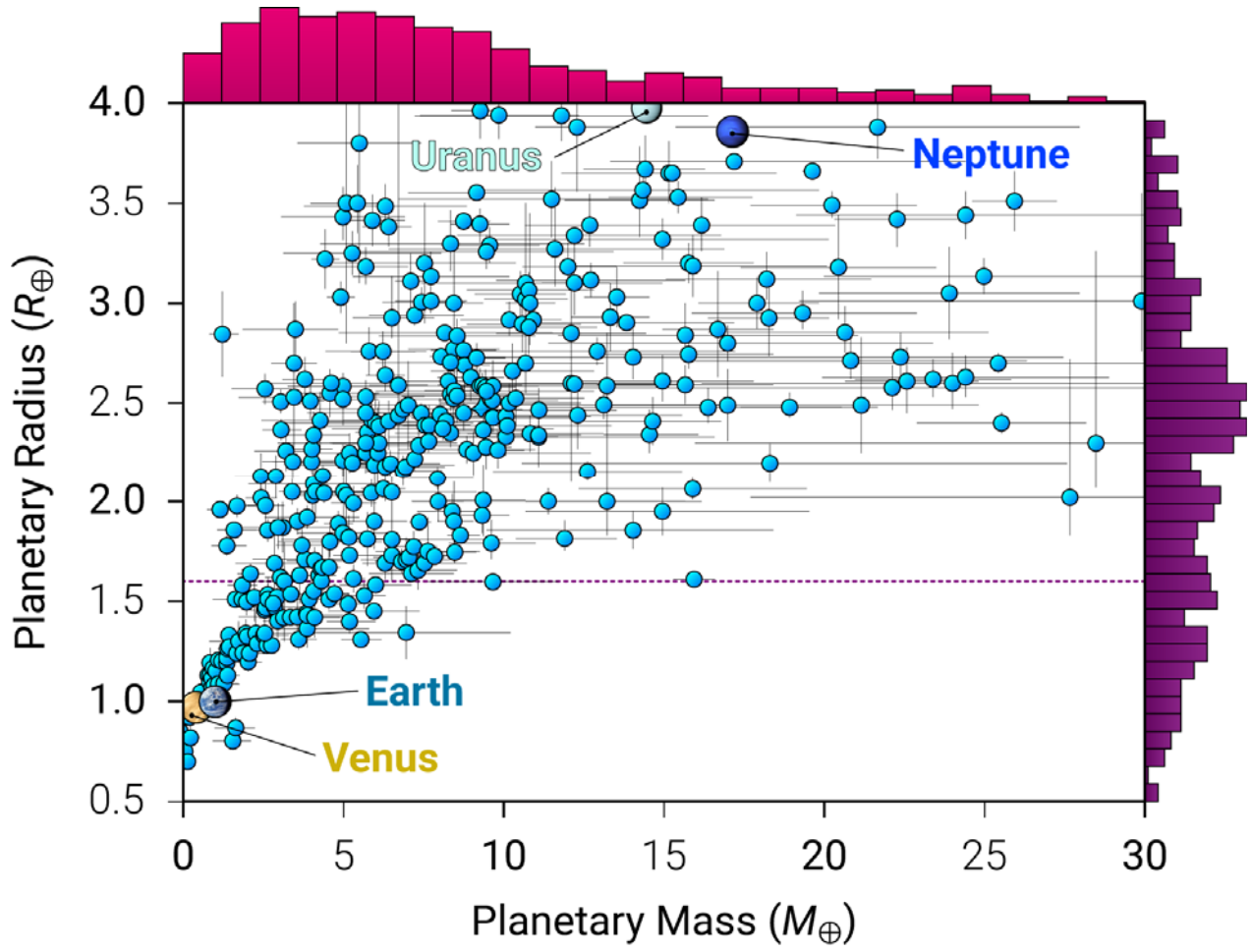

**Fig. 1** A plot of planetary mass versus radius (relative to Earth, often denoted by $\oplus$) for known, confirmed exoplanets as of 4 December 2024. Scatter points (with error bars shown) denote all planets with measured radius $<4\ R_\oplus$ and with measured mass $<30\ M_\oplus$ values ($n = 372$). Projected histograms along the X- and Y-axes show the entire sample of planets with at least that quantity measured ($n = 448$ for mass, $n = 581$ for radius). The dotted red line marks the planetary radius threshold of $<1.6\ R_\oplus$, the rule-of-thumb criterion used to classify planets as "rocky" in current practice. There are dozens of known worlds that meet this criterion, some of which are Earth mass or less and thus may have Solar System cognates. Earth, Venus, Uranus, and Neptune are also shown (but not to scale). Data are from the NASA Exoplanet Archive (exoplanetarchive.ipac.caltech.edu)

and/or orbital period, we must use statistical techniques to account for some going unseen by imperfectly sensitive instruments—such as weighting planet populations by the efficiency with which we can detect them. Complications in extrapolating a parametric occurrence-rate model to longer, Earth-like orbital periods (where lower stellar irradiation mean more temperate climates), for example, include the possibility that close-in rocky exoplanets are overabundant because the high-irradiation environment would cause puffier Neptune-sized planets to lose their atmospheres (Lopez and Rice 2018). Different choices for parametric model fitting can lead to discrepancies of nearly two orders of magnitude in the occurrence rate of Earth-sized planets (Kunimoto and Matthews 2020; Bryson et al. 2021; Bergsten et al. 2022). In general, though, planets smaller than Neptune appear to be tenfold more common than giant planets, with one recent review suggesting that these smaller worlds occur around about one third of Sun-like stars (Winn and Petigura 2024).

How do we know whether an exoplanet, once detected, is "terrestrial"—that is, dominantly rocky, with a metallic core? For most of these worlds, other than orbital period only their radius is measured. For some, we have measured their mass; for precious few, we know both (Fig. 1). Only with mass and radius together can we calculate a planet's bulk density, and thereby might infer with some confidence whether that density is consistent with a mainly rocky and metal composition (e.g., Miozzi et al. 2018). Even then, such analyses can be far from conclusive (as discussed extensively by Baumeister et al. 2025, this collection). Further, for small planets on longer (Earth-like) orbital periods, masses become increasingly difficult to measure by the radial velocity method, and transits become less likely.

Two so-called "Earth twins" have been identified in the current exoplanet census, although the existence of one of them, Kepler-452 b, is questioned (Mullally et al. 2018;

Burke et al. 2019). The other is Kepler-62 f, a planet that orbits a K star, is of radius relative to that of Earth, $R_\oplus$, of 1.46, and receives the equivalent of about half Earth's incident solar radiation (insolation) (Borucki et al. 2013; Weiss et al. 2024). The "Earth twin" label has no robust definition, but generally denotes a planet with radius and orbital period broadly similar to that of Earth and orbiting a host star comparable to our G-class Sun. (See Glaser et al. 2026, this collection for a wider discussion on assessing exoplanet habitability).

Recent estimates of the occurrence rate of these notional twins falls between 0.37 and 0.88 per star (Bryson et al. 2021)—an estimate carrying considerable uncertainty (and optimism) given the necessary extrapolation, but which should improve as future facilities designed to detect a statistically useful yield of small temperate planets around G-class stars come online (e.g., Yahalomi et al. 2023; Kammerer et al. 2022; Stark et al. 2024). In the meantime, highly irradiated rocky planets on close-in orbits, especially around smaller, dimmer M-class stars, are becoming a viable target for spectroscopic characterisation with the space-based NASA/ESA James Webb Space Telescope (JWST) (Lustig-Yaeger et al. submitted, this collection).

Even if we knew the bulk density and orbits of all exoplanets perfectly, deciding on the best terminology with which to classify them is at best ambiguous, and at worst misleading (Tasker et al. 2017). The term "super-Earth" normally refers to a planet below the minimum in the planet radius distribution proposed to separate bodies with and without thick primordial-gas envelopes ($H_2$) (Rogers 2015; Fulton et al. 2017), about $R_\oplus = 1.6$ (see the dotted red line in Fig. 1), without knowing that the planet is otherwise like Earth in bulk composition. (We revisit this point in Sect. 4.2.1.) It is convenient to call these "rocky" or "terrestrial" planets, but we should avoid the temptation to imagine them as blue marbles. Prototypical "rocky" planets may contain substantial mass fractions of volatiles with respect to Earth, such as the TRAPPIST-1 system (Agol et al. 2021; also see Ducrot et al. under review, this collection), whereas others may be relatively volatile-poor, rendering them terra incognitae to a Solar System geoscientist.

Moreover, we are now starting to understand that many sub-Neptune exoplanets may be mainly rock–iron by mass (Kite et al. 2019; Benneke et al. 2024), and even Neptune and Uranus themselves may be principally rocky by mass, hosting large and nominally "terrestrial" planets in their deep interiors, as opposed to the "ice giants" of popular perception (Teanby et al. 2020; Morf and Helled 2025). Although this poorly characterised, fuzzy spread in exoplanet bulk composition may at first suggest a limited use in understanding exoplanets by analogy to processes on worlds known better to us—those comprising the inner Solar System—one might take the countervailing view that topics in Earth and planetary geoscience could apply on a scale much broader than perhaps once believed (e.g., Shorttle et al. 2024).

### 1.2 Current Limitations and Future Abilities

Currently, terrestrial exoplanet characterisation is mostly limited to measurements of orbit, size, and mass. These values have been provided by space-based telescopes such as NASA's Kepler (Borucki 2016), and Transiting Exoplanet Survey Satellite (TESS) (Ricker et al. 2014), which detect exoplanets that transit their host stars along our line of sight. In the near future, ESA's PLAnetary Transits and Oscillations of stars (PLATO) spacecraft will build on those earlier missions by studying much brighter stars (4–11 magnitude) than before. Other ground-based instruments such as the High Accuracy Radial Velocity Planet Searcher (HARPS) spectrograph instrument at the European Southern Observatory employ radial-velocity measurements to determine planetary mass, perhaps most notably in the discovery of Proxima Centauri b (Anglada-Escudé et al. 2016), the nearest (suspected) terrestrial exoplanet.

Yet, although we can now begin to establish population-level statistics for rocky exoplanets, we have some way to go to be able to characterise the surfaces and climates of individual planets (Lagage et al. 2026). Moreover, no terrestrial exoplanet to date has been detected with a temperate atmosphere, and even those large, rocky worlds thought to lie in the Venus zone (Kane et al. 2014) such as TRAPPIST-1 b and c may have thin fluid envelopes or bare-rock surfaces (e.g., Zieba et al. 2023; Ducrot et al. 2025). The exoplanet 55 Cancri e (Demory et al. 2016) is one of the few (relatively) low-mass worlds with a detected atmosphere (Hu et al. 2024); although it is sometimes called a super-Earth, its radius of $R_{\oplus} = 1.8$ exceeds the nominal threshold for that class of body, and its density is inconsistent with a rock–iron composition (instead suggestive of a relatively large mass fraction of low-density material such as carbon).

The major limitation to identifying (hypothesised) temperate atmospheres on transiting exoplanets such as TRAPPIST-1 d and e is the variable luminosity of the most commonly examined type of star, red dwarfs, which makes the observations extremely challenging (e.g., Fauchez et al. 2025; Berardo et al. 2026). The JWST is, in theory, capable of making not only transiting observations but also acquiring reflection (albedo) and emission measurements. JWST reflection spectra (of wavelengths 1–5 µm) may eventually give us insights into the surface albedos of select exoplanets if we can disambiguate those signatures from the effects of any clouds present in their atmospheres. Further, JWST emission spectra via the Mid-Infrared Instrument (MIRI), which operates at ∼5–28 µm, might provide information about surface characteristics for a handful of terrestrial exoplanets. But again, thus far we have no current observations with these space-based instruments that provide conclusive evidence for any terrestrial exoplanet possessing a temperate or even a Venus-like atmosphere.

For ground-based telescopes, the ESO Very Large Telescope's CRIRES+ (Cryogenic high-resolution InfraRed Echelle Spectrograph) instrument is capable of providing high-resolution spectroscopic measurements of exoplanets at 0.95–5.3 µm but, at the time of this writing, no atmospheric observations have been published. In the near future, the ESO Extremely Large Telescope ANDES (ArmazoNes high Dispersion Echelle Spectrograph) instrument will be available with a similar resolution to that of VLT CRIRES+, with a wavelength range of 0.4–1.8 µm and the hope of extending that range to 0.3–2.4 µm, but with a larger light bucket.

There are several space-based concept missions proposed for the longer-term, two notable examples of which include the Large Interferometer For Exoplanets (LIFE: Quanz et al. 2022) and the Habitable Worlds Observatory (HWO: Clery 2023). If built, space-based telescopes such as these would allow for the possibility of characterising the atmosphere of an Earth-size planet orbiting a Sun-like star. This capability is important because of the focus by current technologies—for instance, JWST, VLT, and ELT instruments—on Earth-size exoplanets orbiting M-class red dwarfs (Spohn et al. 2026, this collection). It remains to be determined whether the apparent dearth of terrestrial exoplanets with temperate atmospheres is because of either our present detection limits (e.g., such atmospheres are present but are thin and/or transparent to our detectors) or some intrinsic outcome of terrestrial planetary formation and evolution.

### 1.3 Variations on a Theme?

We are clearly beginning to gain a high-level understanding of the types and frequencies of extrasolar planets—including, increasingly, terrestrial worlds. Routinely investigating the geological and climatological properties of individual rocky exoplanets remains a distant goal, however, and we will likely *never* have the means to visit them with spacecraft.

**Table 1** Key physical characteristics of the inner Solar System worlds. All data are from publicly available sources except for interior structure information, which is taken from Byrne (2020) and references therein, updated with data for Mars from Knapmeyer-Endrun et al. (2021) and Stähler et al. (2021)

| | Mercury | Venus | Earth | Moon | Mars |
|---|---|---|---|---|---|
| Mean Solar Distance (km) | 57,900,000 | 108,200,000 | 149,600,000 | ∼150,000,000 | 227,900,000 |
| Mean Solar Distance (AU) | 0.387 | 0.723 | 1.0 | ∼1.0 | 1.523 |
| Sidereal Orbital Period (Days) | 88 | 225 | 365 | 354 | 687 |
| Sidereal Rotation Period (Days) | 58.6 | 243.0 | 1.0 | 27.3 | 1.0 |
| Obliquity (°) | 0.03 | 177.36[a] | 23.45 | 6.68 | 25.19 |
| Mass (kg) | $3.29 \times 10^{23}$ | $4.87 \times 10^{24}$ | $5.97 \times 10^{24}$ | $7.35 \times 10^{22}$ | $6.39 \times 10^{23}$ |
| Mass ($M_\oplus$) | 0.06 | 0.82 | 1.00 | 0.01 | 0.11 |
| Mean Radius (km) | 2439 | 6052 | 6371 | 1737 | 3390 |
| Mean Radius ($R_\oplus$) | 0.38 | 0.95 | 1.00 | 0.27 | 0.53 |
| Volume (km$^3$) | $6.08 \times 10^{10}$ | $92.85 \times 10^{10}$ | $108.32 \times 10^{10}$ | $2.20 \times 10^{10}$ | $16.32 \times 10^{10}$ |
| Bulk Density (g/cm$^3$) | 5.43 | 5.24 | 5.51 | 3.34 | 3.93 |
| Surface Gravity (m/s$^2$) | 3.70 | 8.87 | 9.81 | 1.63 | 3.71 |
| Core Radius (km) | ∼1900 | ∼3200 | 3449 | 291 | ∼1830 |
| Mantle Thickness (km) | 380 | 2807 | 2874 | 1368 | ∼1512 |
| Crustal Thickness (km) | 35 | 17 | 3/24[b] | 39 | 24–72 |

[a] Venus has a retrograde rotation

[b] Average crustal thickness values for Earth are given for both oceanic (3 km) and continental (24 km) crust

Yet we are not without means to better understand terrestrial planet formation and evolution, for there are numerous such worlds in our own planetary system. After sixty years of exploration, we now have a first-order (or better) understanding of the ways these worlds are similar, and the ways they are different (Table 1). For instance, despite a range in mass of about a factor of ten, we know (or strongly suspect) that Mercury, Venus, Earth, the Moon, and Mars, as well as Io, are all characterised by a silicate crust, a silica–metal mantle, and a metal-rich core that may, in turn, have a molten outer and solid inner portion (Fig. 2). This broad theme has many specific variations, which we describe in this paper, but they are only variations—there is no rocky world in the Solar System so far recognised that is utterly unlike its neighbours.

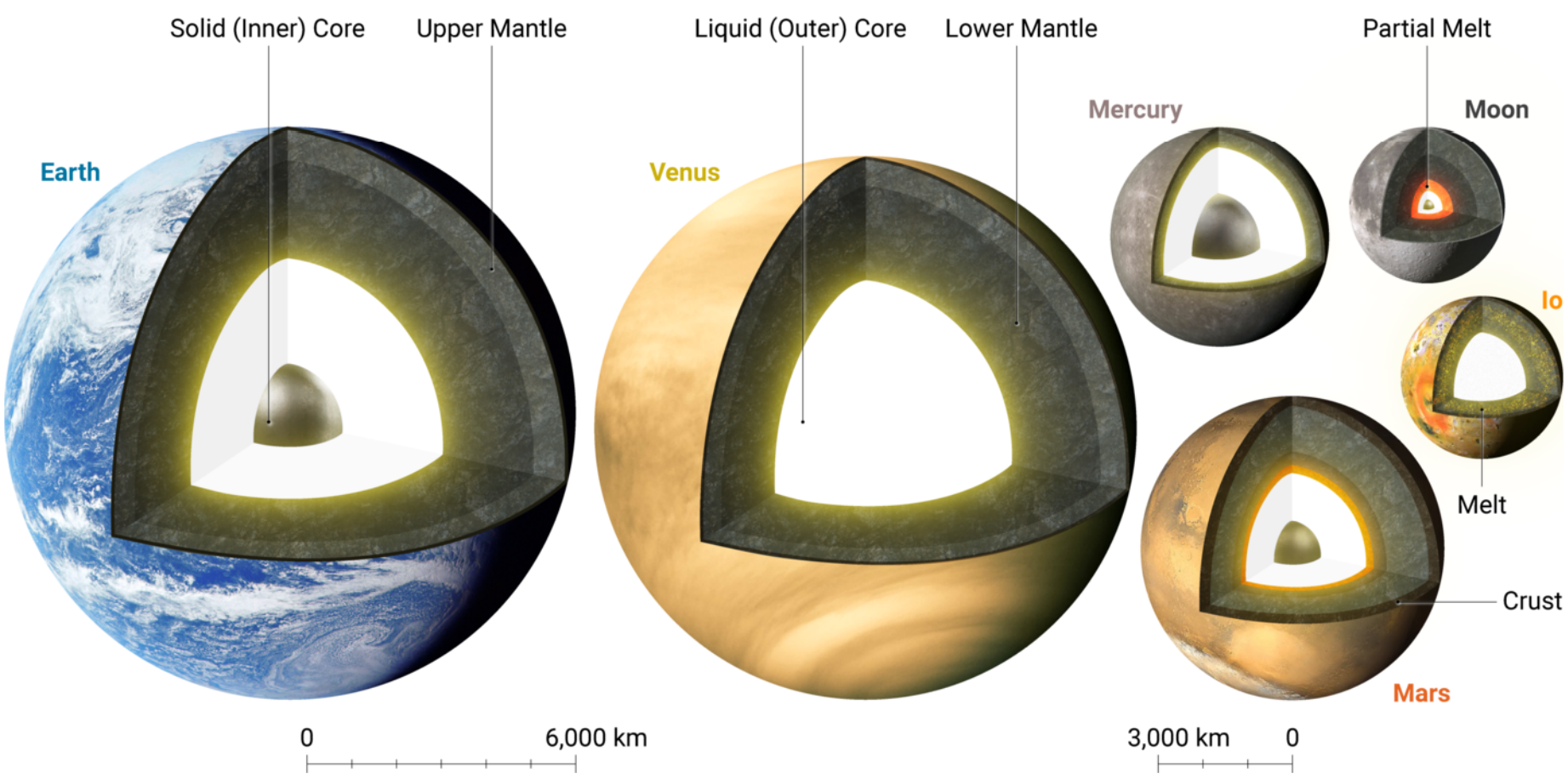

**Fig. 2** Schematic cross sections of the inner Solar System worlds and Io. All bodies and interior structures are to scale (see scale bars at bottom). Major layers are shown, including (where known or suspected) the chemical division between the upper and lower mantle. Despite substantial variations in structure (based in turn on interior pressures and thermal history), the overall theme of ever-greater metal content with depth likely applies to all of these worlds. The interior configurations of Earth, Venus, Mars, Mercury, and the Moon are based on the models described in Byrne (2020) and references therein, together with information on the cores of Mars (Bi et al. 2025) and Mercury (Genova et al. 2019). Note that we have arguably the fewest constraints on a model for Venus' interior structure, including whether the planet possesses a solid inner core, and so the thickness of its component layers are simply scaled relative to those of Earth. The interior of Io is informed by Park et al. (2025)

This fact means that we might reasonably assume that *all* rocky planetary bodies will share *some* basic similarities. By synthesising the major characteristics—geological, climatological, thermal—of the Solar System's terrestrial worlds we have a basis for what to expect of the terrestrial inventories of other planetary systems (even if the particular *arrangements* of those systems differ from ours: Martin and Livio 2015; Gratton et al. 2023). This comparative framework is especially powerful when we consider how the worlds of the Solar System have changed through time.

And so we come to the rationale for this work: to review, compare, and contrast the key properties of the rocky Solar System worlds so as to understand the possible types of planetary outcomes as a function of size, thermal history, and geology. We give overviews of our current understanding of the fundamental properties and phenomena of these bodies, focusing on their interiors, surfaces, and (where present) atmospheres, and with an emphasis on how they have evolved over secular (i.e., geological, or simply long-term) time.

We pay most attention to Earth, since it is the planet about which we know the most—both in its present state, and how that state has come to be. We thus begin Sect. 2 with a summary of the planet in its modern form and then discuss how its interior, surface, and atmosphere have developed over the past several billion years. In Sect. 3, we turn to Venus, Mars, and Mercury and the Moon, before finishing with a brief visit to Jupiter's volcanic moon, Io.

We then in Sect. 4 summarise our present understanding of these worlds, drawing some inferences that might be applied to rocky exoplanets generally, to help guide future efforts to characterise these worlds from a geological (and even habitability) perspective (e.g., Meadows and Barnes 2018; Kane and Byrne 2024). Of course, we know from our surveys (e.g., Fig. 1) that there are types of rocky exoplanets without Solar System counterparts. We there-fore finish Sect. 4 with some examples of what we might be missing in the Solar System, and how such bodies might be similar to, or alien from, what we see in our own backyard.

## 2 Planet Earth

Without question, the planet we understand best is our own. Through thousands of years of living within and interacting with the natural world, and several hundred years of more formalised study of natural sciences, we now have a solid working model of our planet's major characteristics. Thanks to the samples returned to Earth from the Moon since the 1960s, together with hundreds of meteorites known to be lunar in origin, we have gained remarkable insights into how our planet formed that we would not otherwise have had. And over the past few decades especially we better understand how Earth has evolved over Solar System history.

This last point is especially relevant to the study of terrestrial exoplanets—for Earth history has been defined by major changes in its surficial and atmospheric composition, heat generation and loss, insolation, and even by the effects of life itself. To wit, observations of Earth by alien astronomers in billion-year increments would show a different world each time, the only commonality being size, mass, and orbital distance—the very properties we commonly retrieve for exoplanets today.

Here, we review the primary properties of modern Earth—its bulk composition, interior structure, geodynamic regime, major surface features, and its biosphere—and then discuss how key properties of the planet including its volatile abundances and the chemistry of the interior are thought to have changed through time. Beforehand, let us recall that Earth's geological history is divided into four eons that are, in order from oldest to youngest, the Hadean ($>$4.0 billion years ago, or Ga), the Archean (4.0–2.5 Ga), the Proterozoic (2.5–0.54 Ga), and the Phanerozoic (541 million years ago, or Ma, up to the present).

### 2.1 Modern Earth

Earth is a rocky planet with a metallic core, a silicate mantle, and a chemically differentiated crust. It has a solid-body radius of about 6371 km and a mass of $5.97 \times 10^{24}$ kg (Table 1). Today, its surface is largely covered with a global ocean, with broad areas of exposed land and countless smaller islands. The planet is encompassed by an atmosphere that supports the presence of a global biosphere that penetrates at least a few kilometres into the crust (e.g., Drake and Reiners 2021). The modern hypsometry of Earth is bimodal, with two peaks, one at around 4000 meters below sea level and another $\sim$800 meters above sea level (Fig. 3).

Earth's atmosphere comprises nitrogen (78%), oxygen (21%), and argon (0.9%), together with a small amount of other trace gases including carbon dioxide (0.04%) and water vapour (which can vary in abundances of 0–5%). The outermost part of the atmosphere extends to a distance about twice that of the Moon's orbit ($\sim$630,000 km), but over 99% of its mass is contained in the first 100 km (and about three quarters within the first 10 km). Despite being stratified, Earth's atmosphere is a highly dynamic system, with large-scale circulation patterns driven by the uneven distribution of insolation and the Coriolis force, a consequence of the planet's rotation.

Nitrogen and oxygen are the most abundant atmospheric gases, vital to maintaining a climate in which liquid water is stable at the surface. Other trace gases also play an important role in maintaining this climate. For example, carbon dioxide ($CO_2$) and methane are greenhouse gases that trap heat in the atmosphere, whereas ozone ($O_3$) not only shields

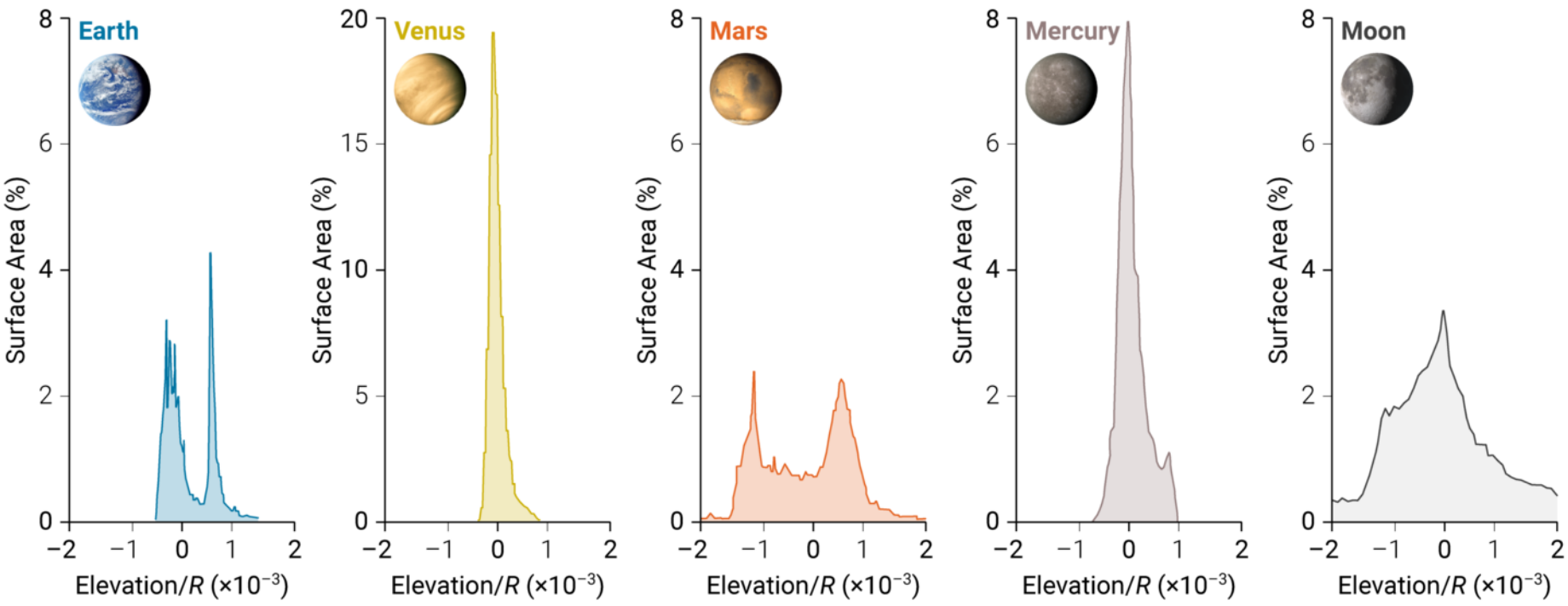

**Fig. 3** Hypsometric curves for the inner Solar System worlds, normalized to their radius (adapted from Criss and Hofmeister 2020, and references therein). Earth and Mars show bimodal hypsometries, attributed respectively to chemical differences in planetary crustal material and to geological (impact or geodynamic) processes. The other three bodies show generally (the Moon) or distinctly (Venus and Mercury) unimodal hypsometries

the surface from harmful solar radiation but dramatically changes the structure of the lower atmosphere and allows for a very dry stratosphere. Water vapour is an especially potent greenhouse gas and a key element of Earth's hydrological cycle, even though its abundance differs depending on location, altitude, and season.

The ocean is mostly composed of water (96%) and dissolved salts (3.5%), with a minor concentration of other dissolved species (including bicarbonate ions, $HCO_3^-$), organic material, and inorganic particles; notably, the ocean is the major carbon reservoir of Earth's atmosphere, and a minor reservoir of dioxygen. Covering 71% of the planet's surface to an average depth of ∼3700 meters (with the greatest depths within long troughs at around 11,000 m), the global ocean contains approximately 97% of all Earth's water (excluding structural water in hydrated minerals in the mantle and non-stoichiometric "water" dissolved in the mantle and core, which together may constitute several oceans' worth). Its volume therefore makes the ocean the largest reservoir of free water on Earth by far, with the remaining water contained in glaciers, ice caps, groundwater, lakes, rivers and as atmospheric vapour. The ocean is a highly dynamic system, with large-scale circulation patterns driven by surface wind and thermohaline circulation. These circulations have a first-order role in redistributing heat and regulating Earth's climate. The global ocean also circulates through seafloor hydrothermal systems on order 10 million years (Myr), which serves to cool the interior and affect the composition of ocean water (Zierenberg et al. 2000).

Although but 29% of the planet's surface is covered by land, the global ocean volume (∼$1.335 \times 10^9$ km$^3$) is merely 0.12% the volume of the entire planet—and so, even if we consider Earth to be an "ocean world", that moniker is literally skin deep. The outer portion of the planet (including that on which the ocean sits) is the crust, of which there are two chemically distinct main types: oceanic and continental (Fig. 4). Oceanic crust is mafic (a portmanteau of "magnesium" and "ferric") in composition, mostly made of basalts and gabbros, which are rich in heavy elements such as iron and magnesium and contain relatively low amounts of silica. This crust is generated at spreading centres in the oceans, where tectonic plates pull apart and the underlying mantle rises, melts from the reduction in pressure, and solidifies as new seafloor. The typical thickness of the oceanic crust is between 5 and 10 kilometres, with a density of about 3000 kg/m$^3$. Oceanic crust is also continuously recycled back into the mantle in subduction zones. For this purely geometric reason, and given the ar-rangements of the continents over the past several hundred million years, there is almost no oceanic crust older than ∼180 million years (Müller et al. 2008) (Fig. 4). Nearly all oceanic crust lies below sea level because of its relatively high density compared with continental crust.

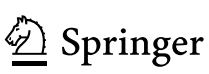

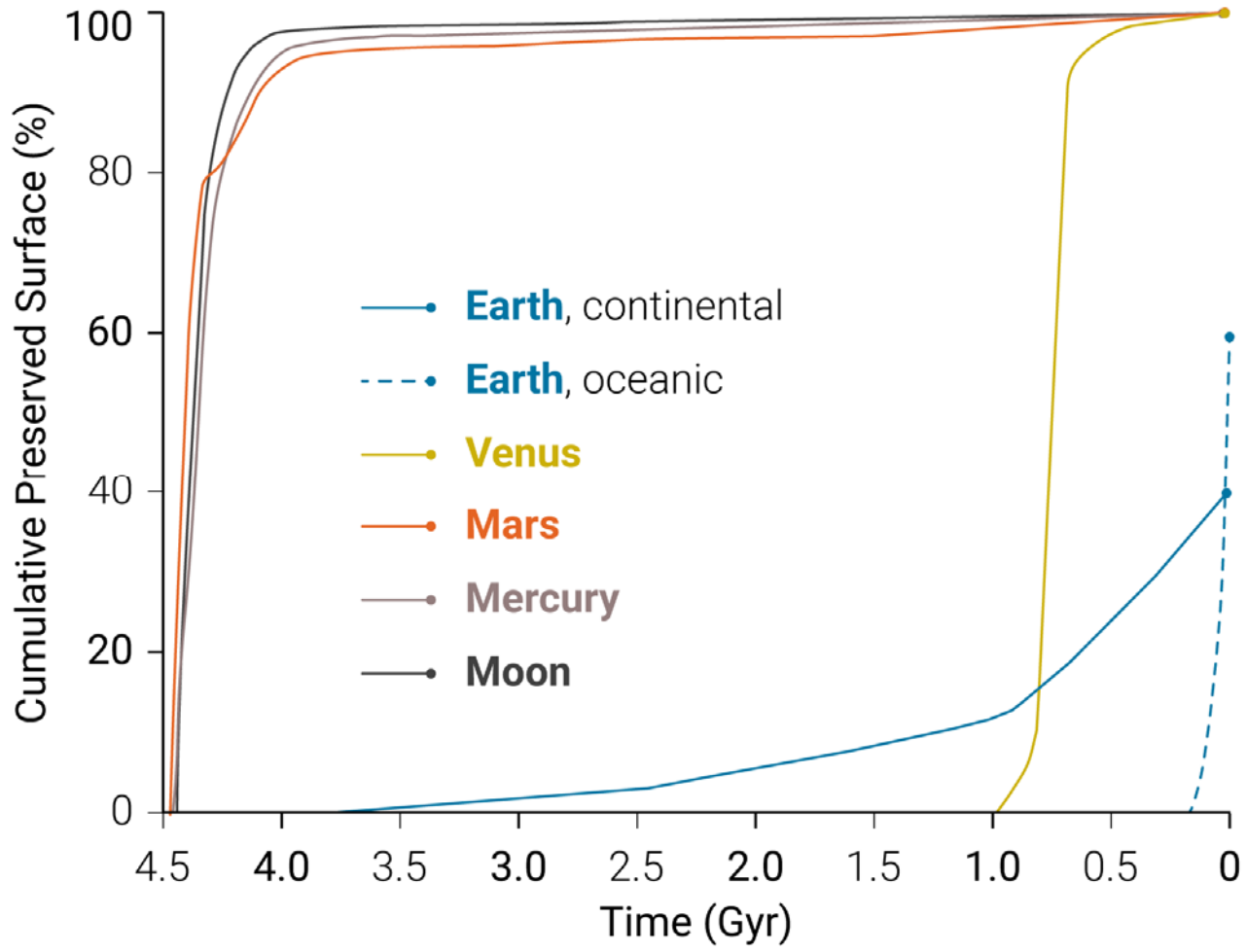

**Fig. 4** Crustal growth rates of the inner Solar System bodies plotted as cumulative crustal mass as a function of time (adapted from McLennan 2022). Unlike its terrestrial neighbours, Earth has two crustal types, with distinct age profiles. Oceanic crust is generally less than ca. 170 Ma (Koppers et al. 2003). For continental crust, the age of preserved Proterozoic and Phanerozoic crust is from Goodwin (1996), with the age of older continental crust extrapolated back to the start of major craton growth at 3.8 Ga (see Sect. 2.3). The history of crustal growth on Venus is essentially unknown, and thus is shown here for an assumed average surface age of around 750 Myr (McKinnon et al. 1997). Photogeological observations indicate that the oldest (post-magma-ocean) crusts of Mars, Mercury, and the Moon were in situ very early, with that of Mars shaped possibly by a giant impact (e.g., Andrews-Hanna et al. 2008) (shown by the small kink in Mars' growth curve) at some point thereafter

Continental crust is thicker, older, and more buoyant than oceanic crust, lying mostly above sea level. With a density of around 2700 kg/m$^3$, its thickness varies between 25 and 90 kilometres. Compositionally, continental crust is more felsic ("feldspar" and "silicic"), mainly made up of granitic rocks that are relatively rich in silica and aluminium and relatively poor in magnesium and iron. Continental crust is much more heterogeneous than oceanic crust. On early Earth, a few billion years ago, continental materials likely formed by the partial melting of hydrated mafic crust (Moyen and Martin 2012; Johnson et al. 2018). On modern Earth, continental crust is mostly created above subduction zones as a result of water leaving subducting slabs, which promotes melting of the overlying mantle wedge and generates highly chemically differentiated, buoyant magmas (Campbell and Taylor 1983). By this argument, liquid water or some other agent able to lower rock melting temperature is essential for the formation of felsic magmas (see Guimond et al. 2026, this collection).

En gros, continents resist subduction and, once formed, tend to stay at the planet's surface. For this reason, continental crust is generally much older than oceanic crust, with rocks reaching ages of at least 4 Gyr (Fig. 4). Continental crust covers only around 40% of Earth's surface (including that portion underwater), but its considerable thickness means it constitutes almost 70% of the total crustal volume.

Whereas the crust (be it oceanic or continental) is defined chemically, and is chemically different than the underlying mantle from whence it came, the outer portion of Earth can be described in mechanical terms, too. We define that part of silicate Earth that is able to sustain geological loads over geological time, and is typified by being relatively cool, strong, and liable to fail through brittle deformation mechanisms, as the lithosphere.

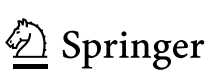

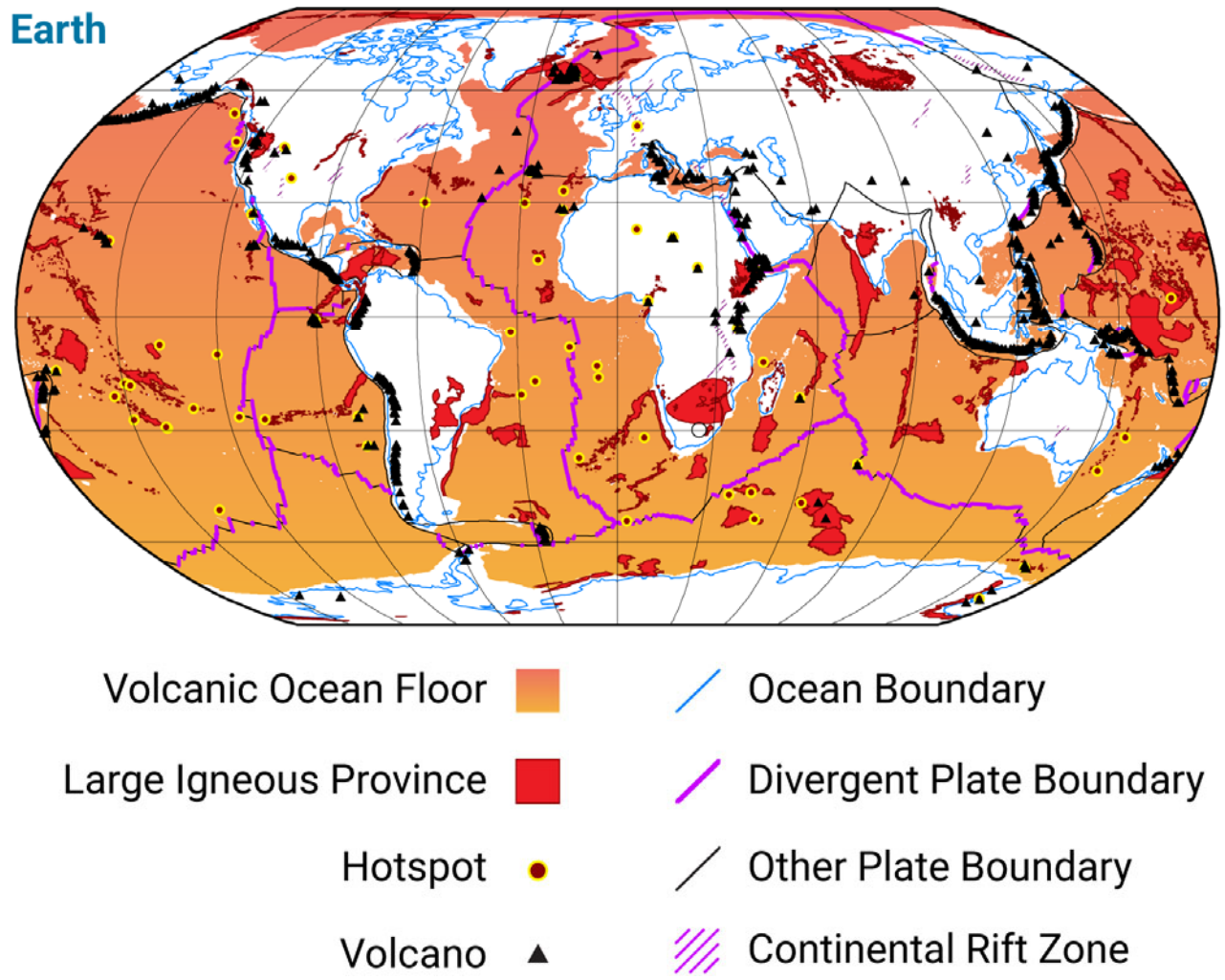

**Fig. 5** A global map of major geological features on Earth. Most of the surface comprises oceanic volcanic plains, with the remainder made of up generally thicker crust that stands above sea level. Much (though not all) of the land is chemically distinct from the oceanic plains. The brittle lithosphere is divided into seven major tectonic plates, themselves delineated by boundaries including spreading centres. Large igneous provinces, composite and shield volcanoes, and rift zones make up the remaining prominent physiographic features on Earth. This and the subsequent maps in Figs. 7–11 are in Robinson projection, centred at 0°E. Adapted from Byrne (2020) and references therein

Earth's lithosphere has an average thickness of 100 km and is composed of both crust (oceanic and continental) and, in most places, the uppermost part of the mantle. Today, the lithosphere is divided into seven major and dozens of smaller rigid plates that move relative to each other over a mechanically weaker layer termed the asthenosphere (Fig. 5). Constantly moving, these plates are mostly driven by the sinking of dense plate margins at subduction zones (Forsyth and Uyeda 1975).

As oceanic lithosphere—made of cold, strong, and brittle oceanic crust and upper mantle—subducts into the deeper, weaker asthenospheric mantle, the crust generated by decompression melting at the spreading centres cools to form new lithosphere to compensate for the loss of subducted material. These cold, sinking slabs also play a role both in the mantle's thermal evolution (see Sect. 2.2, below) and in the planet's rotational inertia. Moreover, a great amount of geological activity occurs along plate boundaries today, such as earthquakes, volcanoes, mountain building, hydrothermal activity, and the formation of mineral deposits (Duarte 2023).

The plate margins where subduction occurs are termed destructive, with subduction itself almost an exaggerated form of thrust faulting. Where new (oceanic) plates form, at constructive margins, normal (extensional) faults accommodate the extension of the crust along extensive rises. And where plates slide past each other, a requirement to geometrically fit on an ellipsoidal planet, their margins are termed transform, after the (often huge) strike-slip faults that define them.

The mantle constitutes around 67% of Earth's mass and is ultramafic, mainly composed of the minerals olivine and pyroxene. Lower in the mantle, olivine transforms into high-pressure phases such as wadsleyite, ringwoodite, and bridgmanite. The mantle is solid, but is capable of flowing at geological time scales because of its rheological properties. It is rheology—how the mantle responds to stress—that accounts for how the uppermost portion is strong and brittle yet the mantle beneath to be geologically weak and capable of convecting. (Below the asthenosphere, ever higher pressures strengthen the mantle again; this lower portion is sometimes referred to as the mesosphere.)

The phenomenon of plate tectonics itself is an expression of mantle convection (e.g., Bercovici et al. 2000; Coltice 2023). On modern Earth, mantle convection is powered by the contrast between internal thermal energy—originating from both primordial and radiogenic sources (Morgan and Vannucchi 2023)—and the planet's persistently cold exterior. The accumulation of thermal energy in the mantle causes it to convect, with the formation of ascending warmer and descending cooler currents. In the present epoch, a substantial part of the mantle convection and surface motion results from the accumulation of cooler surface material that sinks at subduction zones—the subducting slabs of oceanic lithosphere to which we referred above.

The descending material accumulates at the core–mantle boundary, its relatively lower temperature in turn helping to extract thermal energy from the hotter core, which can power huge upwellings (Morgan and Vannucchi 2023). From this perspective, tectonic plates are part of Earth's convective system. However, because the *continental* lithosphere resists being subducted, complex continental interactions occur at the surface, with collisions and splitting of continents and the closing and opening of new oceans in so-called Wilson cycles (Wilson et al. 2019). What's more, from time to time continents seem to cluster together in supercontinents, giving rise to a semi-periodic supercontinent cycle (Rolf et al. 2014).

Beneath the mantle is the core, which today is stratified into a solid inner portion and a liquid outer layer (Fearn and Loper 1981; Hirose et al. 2021). The inner core is a solid iron–nickel sphere with a radius of around 1220 km, and features some additional unknown lighter elements (possibly carbon, silicon, and hydrogen). It has a temperature exceeding 6000 K (e.g., de Wijs et al. 1998), and is subject to tremendous pressure: at least 330 GPa (Alfè et al. 2007). The outer core is $\sim$2200 km thick and is molten, with an estimated temperature of about 3000–4500 K (de Wijs et al. 1998; Anzellini et al. 2013). Currently, inner-core crystallisation of the liquid outer core generates convection currents in the fluid. These convecting patterns create electric currents that, in turn, power Earth's modern dynamo. Although the magnetic field in the dynamo region is complex and multipolar, the dominant component at and above Earth's surface can be described as a dipolar field aligned close to but not exactly with the planet's rotational axis (Roberts and King 2013). The magnetic field produces a comet-shaped magnetosphere about the planet, compressed in the direction facing but extending far into space away from the Sun, which helps shield the atmosphere and surface from the solar wind and cosmic radiation.

The large-scale geodynamic interplay between internal and external processes gives rise to several volatile and nutrient cycles that actively feed back into Earth's geodynamics. For example, water is a powerful enabler of plate tectonics, and its transport to the mantle along descending slabs lowers mantle viscosity, further facilitating mantle convection and plate movement (Lenardic and Seales 2023). Carbon dioxide—which, absent any other effects would increase atmospheric temperature in the long term (see Sect. 3.1)—reacts with rocks to form carbonate minerals through chemical and physical weathering. Those minerals, in turn, are eventually transported to the ocean floor to be returned to the planet interior by way of subduction, before being released back to the atmosphere from arc volcanoes (Walker et al. 1981). The nitrogen (e.g., Förster et al. 2019; Stüeken et al. 2024a) and phosphorus (e.g., Filippelli 2008) cycles are also intrinsically interconnected with several geological processes that make up modern Earth.

A distinct characteristic of Earth is that it has long possessed a biosphere. In the present, most of the planet's biomass is found on land in the form of plants. Life participates in many (bio)geochemical cycles that affect all spheres of Earth's systems. For example, through photosynthesis, life plays a key role in regulating surface temperature and climate and has had a profound effect on the planet's atmospheric composition (Lyons et al. 2014). Life may also have an important role in facilitating plate tectonics (Parnell and Brolly 2021). On the other hand, the surface motion of continents promotes speciation and biological evolution (e.g., Lourenço et al. 2026, this collection). Indeed, it can be said that life is co-evolving with the planet's geodynamic progression (Stern and Gerya 2023).

## 2.2 Secular Evolution of the Interior

Earth is thought to have formed through a process in which increasingly larger planetesimals collided to form the nascent planet, heating it beyond the melting temperature of its rocky components. This heating resulted in the formation of a global magma ocean and the differentiation of the interior, which led to the development of an iron-rich, molten core and a comparatively lighter silicate mantle. The cooling and crystallisation of the surficial magma ocean and the degassing of the interior established the primordial crust and the planet's first true atmosphere. Because of these high interior (and possibly surface) temperatures, the outer proto-lithosphere was likely quite thin, and volcanic activity was probably widespread.

At present, Earth is losing heat at a rate of $\sim$41 TW (Lucazeau 2019), the majority of which is from secular cooling from the planet's formation. The mantle is thought to have cooled by about 200–300 K since the Hadean eon, with the core cooling by a similar amount (see Davies et al. 2015 for a review). However, it is possible that this cooling has been episodic rather than steady, in which case fluctuations would superimpose upon the secular cooling trend (cf. the model incorporating continental growth of Walzer and Hendel 2022). Even so, the overall pattern of secular mantle cooling is thought to have led to profound changes in Earth's tectonic modes, amounts of interior melting, and even melt composition (Herzberg et al. 2010). Komatiites, for example—rocks known for their high Mg content—are interpreted to be a consequence of higher degrees of partial melting, and are much more commonly found in Archean-age settings than in younger rocks (e.g., Nisbet et al. 1993).

The path taken by Earth to the modern plate tectonics regime was gradual and probably involved a warmer, more deformable lithosphere than today (e.g., Cawood et al. 2022; Chowdhury et al. 2025). At least some of this crust was likely recycled through delamination and dripping (Johnson et al. 2014). Nevertheless, geological evidence also points to at least portions of the crust being tectonically immobile at first (Lowe 2025), before showing evidence for substantial lateral shortening (Heubeck et al. 2023). Impacts likely played an important role on the very early Earth, as well (Marchi et al. 2014). This transitional mode has been termed "squishy-lid" tectonics (Rozel et al. 2017; Lourenço et al. 2020), and is characterised by substantial volumes of shallow intrusive bodies that serve to reduce the thickness of the lithosphere. This process may be occurring on Venus today (Rolf et al. 2022), at least in the planet's lowlands (Byrne et al. 2021) (see Sect. 3.1). Stagnant- and squishy-lid regimes may have overlapped in time on Earth, operating in different parts of the planet.

The cooling of the core is governed by heat transfer through the mantle. Christensen (2010) reviewed the scaling of numerical and laboratory dynamos and concluded that the magnetic field strength is proportionate to the radius of the core and the (thermal and/or chemical) buoyancy flux. When Earth's interior magnetic field arose remains debated; some studies are suggestive of a Hadean geodynamo (e.g., Tarduno et al. 2015) but this view is contested, with others finding that the dynamo may date to the mid-Archean (Borlina et al. 2020) or even to the early Archean (Nichols et al. 2024). As for the inner core itself, estimates for when it began to develop range from around 2 Ga or even earlier (Smirnov et al. 2011), to about 1 Ga (Labrosse et al. 2001), to perhaps even the late Proterozoic (Bono et al. 2019).

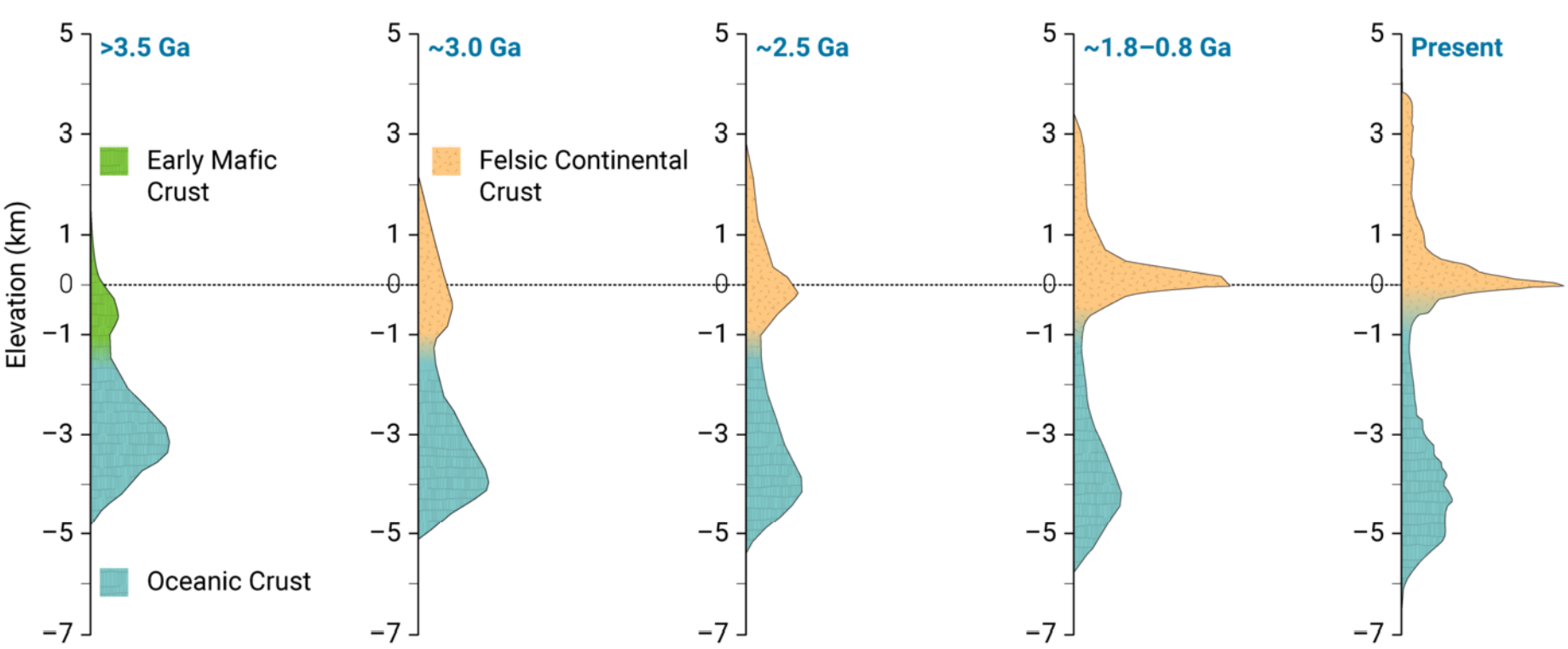

**Fig. 6** A conceptual model for how Earth's hypsometric curve has grown, with schematic curves for the early, mid-, and late Archean, and the late Paleoproterozoic to early Neoproterozoic, all highlighting the progressive development of its bimodal hypsometry (from Cawood and Hawkesworth 2019). The horizontal axis represents proportional area, and the horizontal dashed line demarcates modern sea level. In the Hadean and early Archean, Earth likely had little exposed land

## 2.3 Secular Evolution of the Solid Surface

The nature, amount, and rate of production of crustal materials have changed throughout Earth history, even if the entire surface of the planet has been covered in *some* type of crust since its formation. Moreover, crustal growth rates likely varied hand-in-hand with changes in tectonic regime, both as a consequence and driver of those secular changes. However, the limited and fragmented preserved rock record of early Earth has led to the formulation of numerous models for the rate of crustal growth and the proportion of continental crust, as well as the tectonic modes that drove these processes. For example, although a range of ages has been proposed for the onset of plate tectonics (e.g., see the discussion in Korenaga 2013), we consider that, on the basis of the available geological record (e.g., Cawood et al. 2025), plate tectonics *at least as it operates and is expressed today* has not been the modus operandi for the cooling of the mantle since Earth condensed from the protoplanetary cloud.

In Fig. 6, we show a schematic model for how Earth's crustal volume (and, by turn, hypsometry) may have changed with time. This figure applies the succession of Earth tectonic modes outlined by Cawood et al. (2022), showing first the growth and recycling of an initial crust resulting from the solidification of Earth's final magma ocean, then an early Hadean crust formed under a (hot) stagnant-lid regime (cf. Sect. 3.1), followed by the main phase of craton formation involving bimodal mafic and felsic crust (inferred to have developed under a squishy-lid regime), and finally the oceanic and continental crustal growth records associated with onset of the plate tectonics mode (Guimond et al. 2026, this collection).

The first global solid surface on Earth likely formed post-accretion during solidification of the planet's final magma ocean (cf. Elkins-Tanton 2012), which presumably followed the Moon-forming impact (Canup and Asphaug 2001). This event, in which the Moon is thought to have resulted from a giant impact between proto-Earth and a Mars-sized object called Theia (Hartmann and Davis 1975; Benz et al. 1986; Canup and Asphaug 2001), along with other major impact events during the first few tens of millions of years, must surely have resurfaced or even destroyed whatever crust had been present before, which was possibly a mafic or ultramafic quench crust. The absence of any recognized areas of preserved initial crust on Mars, Mercury, or the Moon—that is, areas of the surface with crater retention records corresponding to absolute model ages of about that of the Solar System itself—suggests that the initial crust was rapidly recycled or even destroyed during the first few tens of millions of years of Earth history (Mojzsis et al. 2019). A comparison of the preserved crustal record of these other bodies with that of Earth (Byrne 2020) further indicates that replacement of the magma ocean-generated crust by the first true crust (or "secondary" crust, using the parlance of Taylor (1989)) was rapid, and presumably largely mafic. This crust would have been considerably thicker than modern oceanic crust, of the order of 20 km based on substantially higher mantle potential temperatures resulting in greater amounts of partial melting than today (Sleep and Windley 1982; Herzberg et al. 2010).

There is no evidence that this mafic crust necessarily formed in a manner similar to modern mafic oceanic crust at spreading centres, but it may have been emplaced akin to the vast basaltic lava plains of the other inner Solar System worlds (Byrne 2020), albeit underwater (Flament et al. 2008; Cawood et al. 2022). This primordial crust is thought to have included felsic components to account for the Hadean zircons (e.g., Harrison 2009), although Whitehouse et al. (2017) argued that these zircons were likely derived from mafic materials. Some felsic crust is preserved in the late Hadean to earliest Archean Acasta Gneiss—where felsic melts are inferred to have formed by shallow magmatic processes including assimilation of hydrated rocks (Reimink et al. 2014) or re-melting (Johnson et al. 2018) of mafic to ultramafic crust (cf. Mojzsis et al. 2014). Nonetheless, these felsic rocks were likely not produced at the volume of later continental crust, given their relative dearth in the rock record. Ongoing igneous activity throughout the remainder of the Hadean would have been therefore dominantly mafic, extrusive, voluminous in nature, and driven by overturn events in response to high mantle potential temperature, probably enhanced by impact bombardment (e.g., Moresi and Solomatov 1998; O'Neill et al. 2017).

The rock record indicates that the formation of preserved Archean cratons (continental cores) commenced around 3.8 Ga, which incorporated and reworked remnants of the Hadean crust (Cawood et al. 2022; Vervoort and Kemp 2025). There is little evidence for Hadean crust being preserved into the Proterozoic; much of our understanding of the Hadean comes from zircon crystals incorporated into later rocks (Dunn et al. 2005). Thus, on Fig. 6 we suggest recycling of the earlier Hadean crust commenced around 3.8 Ga, corresponding with the start of the main pulse of craton formation, and was complete by 2.5 Ga with the end of craton stabilisation. Craton formation appears to have involved an initial phase of growth extending to around 3.2 Ga, followed by a stabilisation phase that was generally of order a few tens of millions of years but occurred on different cratons at different times between 3.2 and 2.5 Ga (Laurent et al. 2014; Cawood et al. 2018). We interpret this timeframe as corresponding to the transition from broad zones of deformation, marked by widespread melting of the lower crust within a squishy lid, to more focused zones of deformation. This later deformation may have initially been associated with localised and short-lived regions of convergence and divergence of increasingly rigid and, likely, larger crustal blocks as they began to assume the characteristics of modern tectonic plates.

By the end of the Archean, stabilization of the cratons is taken to mark the start of broad-scale, rigid, mobile-lid tectonics—in other words, the onset of the phenomenon we know as plate tectonics (Cawood et al. 2025), a process likely helped by (if not necessarily *requiring*) abundant liquid water (Bolfan-Casanova 2007; Korenaga 2020). Crustal growth during the craton-building phase included the cratonic granite greenstone association along with the formation of early mafic "oceanic" crust. There was likely a compositional continuum between the preserved greenstone crust and the early oceanic crust, the latter of which was recycled. The felsic component of the cratonic crust formed through melting of thickened mafic crust (whether termed early oceanic or cratonic greenstone) (Moyen and Martin 2012). A variety of criteria has been invoked to compare greenstone belts, including those at Isua, Greenland, to ophiolites and so infer that they represent ancient oceanic lithosphere (e.g., Komiya et al. 1999; Furnes and Dilek 2022). Geochemical data, however, suggest that the mafic–ultramafic associations preserved within the craton blocks are distinct from Proterozoic or Phanerozoic ophiolites (upthrust slices of mantle), and the majority thus did not form at sites such as mid-ocean ridges. Instead, these lithologies may represent mantle plumes in a rifted continental setting (e.g., Brown et al. 2024).

The inferred increasing rigidity and stabilisation of cratonic lithosphere post-3.2 Ga is taken as a necessary precondition for the initiation of the mobile-lid plate tectonic mode, although this phenomenon did not become global in extent until the beginning of the Proterozoic. In Fig. 6, we therefore take 3.2 Ga as the start of the generation of oceanic and continental lithosphere by plate tectonics and the resultant bimodal hypsometry that characterises modern Earth (Fig. 3). Craton formation (3.8–3.2 Ga) with associated felsic crust generation likely led to a broadening of the initial quasi-unimodal hypsometry of early Earth, prior to the post-Archean development of a pronounced bimodal hypsometry. Plate tectonics led to widespread recycling of the early mafic crust. At least some cratonic crust was of sufficient thickness and stability to resist this recycling, but the truncation of modern craton boundaries by Proterozoic and younger structures indicates that there was at least some reworking and loss of cratonic lithosphere to the planet interior.

Most models of continental growth give amounts of 60–70% of the present continental crust at the end of the Archean, yet the crust of this age constitutes less than 10% of current area—such that we must conclude that reworking and recycling of Archean crust has continued to the present. In our conceptual diagram (Fig. 4), we suggest that early oceanic and cratonic crust would start to be destroyed with the start of plate tectonics at perhaps 3.2 Ga, with the former completely recycled by the end of the Archean. For continental growth, we use the curve of Dhuime et al. (2012), for which the percent of continental crust increases from $\sim$30% at 2.5 Ga to 40% today. Thus, the proportion of oceanic crust at 2.5 Ga is around 70% of the planetary total, with about 60% having formed at the linear, mid-ocean ridge spreading system that had by then developed, and the remaining 10% or less constituting the early mafic crust that occupied the oceans outboard of cratons. Models invoking higher proportions of early continental crust than Dhuime et al. (2012) would have a corresponding lower proportion of oceanic crust, or could have recycled more continental crust over time (cf. Rosas and Korenaga 2018; Höning and Spohn 2023; Guimond et al. 2026, this collection). Furthermore, $^{142}$Nd isotopic systems suggest that Hadean crust was largely recycled into the mantle by the end of the Archean (e.g., Caro 2011).

## 2.4 Secular Evolution of the Atmosphere and Climate

As for the interior and surface, Earth's atmosphere has undergone substantial compositional and perhaps even volumetric changes since the planet first formed. For example, the realisation decades ago that the Sun's luminosity has increased by nearly 30% since first entering the main sequence more than 4.4 Ga (Schwarzschild 1958; Gough 1981) gave rise to the concept of the "Faint Young Sun Paradox" (e.g., Ringwood 1961; Sagan and Mullen 1972). The original concept held that, if Earth's atmospheric composition and density were the same 3.8–3.6 Ga as today, then the planet would have existed in a climatic "snowball" state as there would have been insufficient greenhouse gases to keep the surface temperate under this reduced insolation.

The paradox arises, however, because various proxy data analysed over the past few decades (e.g., Wilde et al. 2001; Mojzsis et al. 2001; Valley et al. 2002) have shown that the planet was likely temperate from early in its history. A variety of greenhouse gases has been proposed to compensate for the Sun's lower luminosity, with current thinking being that larger partial pressures of $CO_2$ provide enough warming for a modern, $\sim$1-bar atmospheric pressure (e.g., Charnay et al. 2017; Feulner et al. 2023). A study by Som et al. (2012) proposed that features interpreted as 2.7 Gyr-old fossil raindrop imprints correlate with an atmospheric pressure range of 0.5–1.1 bars, although Kavanagh and Goldblatt (2015) found the likely value to be at the upper end of this range. Work by Marty et al. (2013) with N and Ar isotopes from fluid inclusions trapped in 3–3.5 Gyr-old quartz returned very similar atmospheric pressure values. Later, Som et al. (2016) analysed gas bubbles in 2.7 Gyr-old basaltic lava flows and found a surface pressure of $0.23 \pm 0.23$ bar that, when combined with those earlier studies, gives an upper limit of 0.5 bar for late-Archean atmospheric pressure. Payne et al. (2020) came to a similar conclusion by analysing micrometeorites.

In any case, it is clear given prevailing temperate conditions and surface liquid water availability that the late Hadean and early Archean were capable of *supporting* life (as we know it), if not necessarily *hosting* it. From the time of the post-Apollo period until recently it was widely believed that the "Cool Early Earth" (Valley et al. 2002) at 4–4.4 Ga was followed by a "Late Heavy Bombardment" that would have sterilized the planet from about 4 Ga to some time around 3.8 Ga. However, more recent work better supports an *early* heavy bombardment (see the discussions in, for example, Brasser et al. (2016), Morbidelli et al. (2018), and Mojzsis et al. (2019)), which may better conform with available evidence for a life-supporting environment throughout these early epochs, even if continued albeit punctuated bombardment affected the planet's habitability (e.g., Marchi et al. 2021).

It is a different question of when life actually arose within this potentially habitable environment (e.g., Westall et al. 2023; Stüeken et al. 2024b). The earliest potential evidence for early life on Earth is from interpreted microfossils from around 3.8 Ga, if not before (e.g., Dodd et al. 2017), although stronger evidence exists from morphological and geochemical biosignatures covering the Paleoarchean at 3.6–3.2 Ga (cf. Homann 2019). Perhaps the earliest biochemistry utilized the acetyl-coenzyme-A (CoA) pathway to fix $CO_2$ in deep marine hydrothermal settings (Russell and Martin 2004; Preiner et al. 2020). Later, among the most prolific lifeforms in the Archean were likely methanogenic microbes that produced $CH_4$ from the reaction of $CO_2 + 4H_2 \rightarrow CH_4 + 2H_2O$, where the $H_2$ and $CO_2$ would have been supplied by volcanic outgassing. Kasting and Catling (2003) speculated that these archaea could even have produced sufficient amounts of $CH_4$ in the Archean atmosphere to substantially affect the climate (e.g., Kasting and Catling 2003). At present, $CH_4$ abundances in the Archean remain uncertain (Catling and Zahnle 2020).

One important ingredient for life is climate stability, and over the long term such stability requires some form of volatile cycling. As we have discussed, early Earth was certainly not in the same tectonic and climate state it is today, yet the picture of the planet in this time is incomplete (Pinti 2025). The amount of exposed land in the late Hadean and early Archean was likely far less than today. Although unconformities, clastic sedimentary sequences, and palaeosols demonstrate that there was exposed land since at least 3.5 Ga (and possibly as far back as 3.8 Ga), geochemical proxies and the submarine emplacement of continental volcanism point to very limited subaerial exposure of rocks (Flament et al. 2008; Cawood et al. 2022; Chowdhury et al. 2025). In this context, the ample evidence for a temperate climate in the early Archean, notably the presence of surface liquid water, is striking—implying some form of efficient volatile cycling, even if we do not yet have a firm idea of exactly how it operated.

The transition from the anoxic Hadean and Archean to the Proterozoic around 2.4 Ga is defined by the rise of substantial amounts of molecular oxygen in the atmosphere that was likely from the combination of oxygen production by cyanobacteria and by the efficient burial of organic matter (Kaufman et al. 2007; Papineau et al. 2007; Olejarz et al. 2021). The oxygenation of the planet's oceans largely followed that of the atmosphere, although where and how much oxygen the oceans hold varied considerably in space and time (Reinhard and Planavsky 2022). This period also coincides with the first documented "snowball" Earth (e.g., Roscoe 1969; Kirschvink 1992; Prasad and Roscoe 1996). Snowball Earth's lower temperatures could have been precipitated by a drop in partial pressure of the methanogenically produced $CH_4$ that can be expected with higher $O_2$ abundances (since $CH_4$ is quickly destroyed in reactions with OH radicals from $H_2O$ photolysis). As $CH_4$ is a potent greenhouse gas, its demise would have greatly affected surface temperatures; in contrast, an increase in the radiatively inactive $O_2$ would, like $N_2$, have had relatively less effect on climate.

The final major transition period in Earth's history takes place around the boundary of the Proterozoic and Phanerozoic (541 Ma), when complex life geologically rapidly expanded in form and environmental niche in a radiation that started in the late Neoproterozoic (Servais et al. 2023). Since biostratigraphy can most readily be practised for Phanerozoic rocks, we know of at least five major mass-extinction events within the last half-billion years. One of these extinctions can be confidently attributed to exterior influences—the asteroid strike that in killing the non-avian dinosaurs marked the Cretaceous–Palaeogene boundary at 66 Ma—with the rest being endogenic in nature.

But certainly the most severe loss of species diversity in Earth history, at the Permian–Triassic boundary (252 Ma), is principally attributed to sudden and substantial injections of $CO_2$ and sulphur dioxide ($SO_2$) into the atmosphere from large igneous province (LIP) eruptions (e.g., Bond and Wignall 2014). This extinction event saw the loss of $\sim$90% of all land species and $\sim$80% of those in the oceans (e.g., Retallack 1995; Benton 2015). This extinction was caused by a cascade of events (warming, ocean anoxia, ocean acidification) related principally to large-scale volcanism associated with the Siberian Traps in what is today Russia (e.g., Wignall 2001; Svensen et al. 2009; Black et al. 2018), which pumped 1–10 teratonnes of $CO_2$ into the atmosphere (e.g., Wu et al. 2021). The exact cause(s) of LIP volcanism remain under debate, with one leading mechanism the genesis of mantle plumes at the borders of the large, low-velocity shear provinces above the core–mantle boundary (Garnero et al. 2007; Kreielkamp et al. 2022), even if these features remain controversial as an LIP driver (e.g., Davies et al. 2015; Doubrovine et al. 2016). Whatever their cause, LIPs demonstrate unequivocally that, in addition to the very long-term changes to Earth's atmospheric composition since formation, ongoing geological activity can have profound if relatively short-lived effects on that composition.

## 2.5 The Different Faces of Earth

As we review here, Earth has witnessed major changes in its internal thermal state, water–land ratio, atmospheric composition, and perhaps even its atmospheric density throughout its history. The late Hadean/early Archean was likely characterized by a mostly water-covered world with an $N_2$-dominated atmosphere and with $CO_2$ partial pressures of 10–30% (e.g., Charnay et al. 2017; Feulner et al. 2023). $CH_4$ may have also been a component at less than 1%, but its exact abundance remains debated.

As Earth transitioned from the Archean to the Proterozoic, $CO_2$ partial pressures began to drop as $O_2$ partial pressures increased considerably. At the same time, the amount of land exposed to the atmosphere increased markedly, leading to increased erosion, runoff, and nutrient cycling to oceans, and with mountain chains coming to demarcate plate boundaries and joining that topography formed earlier by impact bombardment and volcanic activity.

All of these changes through Earth history have had major impacts on the planet's surface, planetary albedo, weather patterns, and ultimately the sorts of observables that exoplanet researchers will rely upon to characterize the climates of Earth-size worlds around other stars (Westall 2024; Westall et al. 2025). Indeed, the very *colour* of Earth may have changed through time, perhaps from an early orange through to its familiar blue of today, as a function of atmospheric composition, surface temperature, and haze abundances (e.g., Arney et al. 2016). These secular changes in our own world tell us that we should be mindful of how the appearance of large rocky exoworlds may vary throughout their own geological evolution. And, given that life has been present on Earth for at least $\sim$3.5 Gyr, we should remember that a living world need not necessarily look the way our one does today.

## 3 The Other Rocky Worlds in the Solar System

### 3.1 Venus

To first order, there is very little to distinguish Venus from Earth. With a mass of 0.82 $M_\oplus$ and a radius of 0.95 $R_\oplus$ (Table 1), both worlds are similar in size, mass, and (presumably) bulk composition. Venus orbits the Sun closer than does Earth, receiving about twice as much insolation, but still within the star's habitable zone. Both planets have atmospheres. From afar, the Solar System hosts two large rocky planets that may be habitable.

On the surface, Venus hosts an array of familiar geological features (Fig. 7). About four-fifths of the surface comprises low-lying plains, much of them covered in lavas (Ivanov and Head 2013) that are, in the main, basaltic (Surkov et al. 1983). Scattered across these plains are volcanoes of varying sizes, from giant shield volcanoes (more voluminous than, if not as tall as, their Martian peers: see next section) to smaller volcanic cones (e.g., Head et al. 1991) that collectively number in the tens of thousands (Hahn and Byrne 2023). Extensive and topographically elevated rift zones, morphologically akin to those in continental settings on Earth, circle much of the low latitudes (Solomon et al. 1991; Ivanov and Head 2011). The remainder of the surface physiography is defined by highlands that show substantial tectonic deformation, termed "tesserae" (Barsukov et al. 1986); these highlands have morphological and geophysical properties that have led to the suggestion that they are Venus' counterparts to the continents on Earth (e.g., Hashimoto et al. 2008).

Yet that is where the major similarities end. The surface conditions at Venus, measured first remotely and then by in situ missions, are an astonishing 750 K and 92 standard atmospheres (Marov et al. 1973). The air is 93.5% $CO_2$ with negligible water vapour (Vinogradov et al. 1968), and the surface is shrouded by a global layer of sulphuric acid clouds that extend from an altitude of about 47 km to around 62 km (e.g., Avduevsky et al. 1973). Intriguingly, the Venus atmosphere was found to have a deuterium/hydrogen (D/H) ratio around a factor

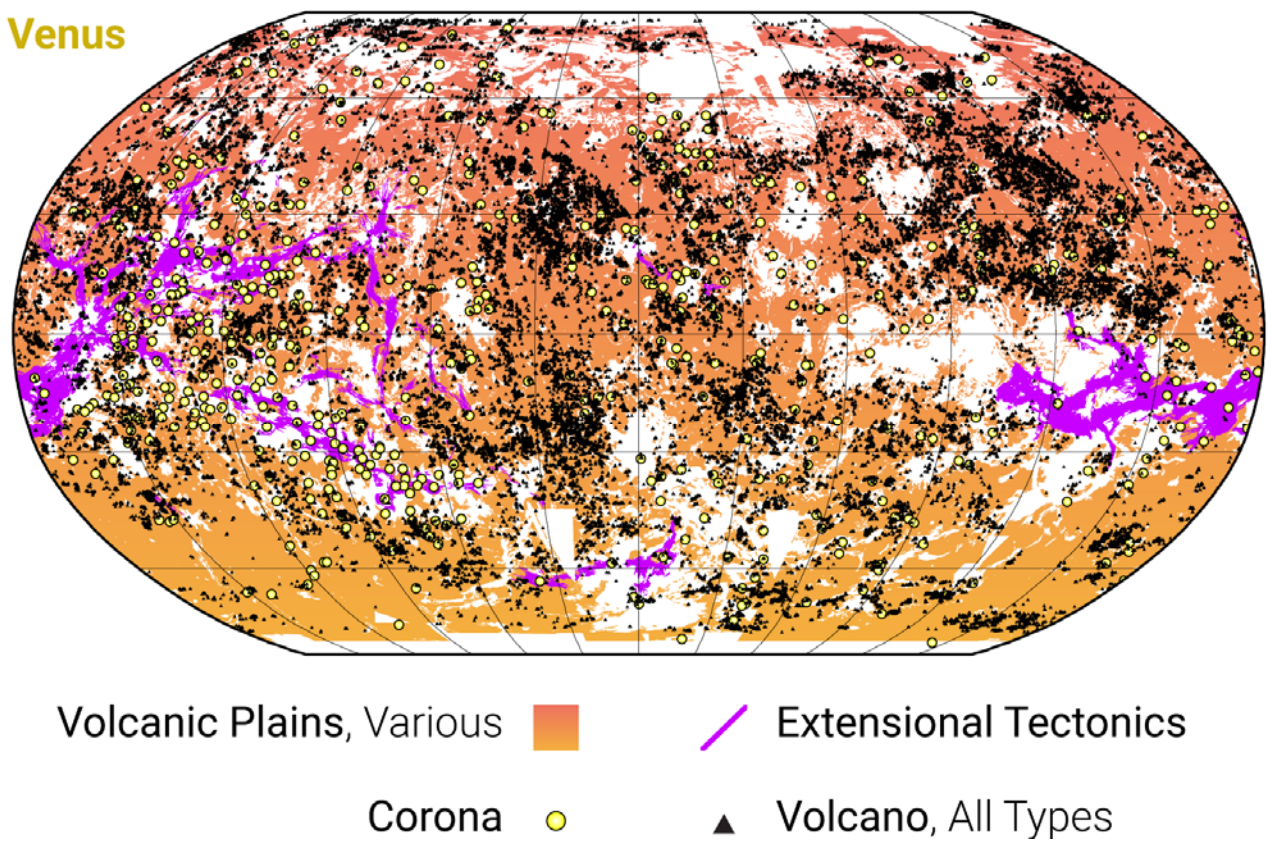

**Fig. 7** A global map of major geological features on Venus. About four fifths of the surface are low-lying volcanic plains, with the remainder consisting of highly tectonically deformed highlands and expansive rift zones. Magma upwellings (the "coronae") abound, and the planet is positively festooned with volcanic edifices, the vast majority of which are situated in the plains. Available remotely sensed data indicate that modern Venus shows no evidence for Earth-style plate tectonic boundaries. Adapted from Byrne (2020) and references therein, with the volcanic edifices from the catalogue of Hahn and Byrne (2023)

of 100 greater than that of Earth, a value that, although subject to considerable measurement uncertainty, was interpreted to indicate the substantial loss of water at some point in Venus' past (Donahue et al. 1982). Modern Venus is regarded as existing in a post-runaway greenhouse state.

The heavily tectonically deformed highlands notwithstanding, the planet shows no evidence for Earth oceanic-style plate tectonics as defined by a global network of rigid plates demarcated by lengthy spreading centres, vast subduction troughs, and transform faults (e.g., Solomon et al. 1991). There are localised subduction systems associated with some of the coronae (McKenzie et al. 1992; Sandwell and Schubert 1992; Schubert and Sandwell 1995; Davaille et al. 2017)—large, quasi-circular volcanotectonic landforms thought to reflect mantle upwellings (Stofan et al. 1992)—and certainly there has been much extension of the crust as manifest by the rift zones. But the morphological signature of plate tectonics is missing on Venus. Notably, the planet's hypsometric curve is strongly unimodal (Fig. 3).

That is not to say that the planet is geologically dead. The clouds are not photochemically stable over geological time, and must be replenished (although the rate of which is debated: Fegley and Prinn 1989; Bullock and Grinspoon 2001), presumably by ongoing volcanism. Such volcanic activity was suspected on the basis of anomalous thermal emissivity measurements from the ESA Venus Express spacecraft (Smrekar et al. 2010), and later confirmed by the discovery of changes to volcanic landscapes during the Magellan mission itself (Herrick and Hensley 2023; Sulcanese et al. 2024). There is extensive tectonic deformation of the planet's low-lying volcanic plains (Solomon et al. 1991) that points to considerable internal activity, including the jostling of some crustal blocks in the geologically recent in a manner akin to pack ice, perhaps driven by mantle motion (Byrne et al. 2021).

The nature of this tectonic deformation has motivated considerable interest in Venus' modern geodynamic state. Whereas Mars, Mercury, and the Moon are entirely consistent with the so-called "stagnant lid" geodynamic regime—under which interior cooling is predominantly accomplished by conduction through a mechanically strong, rigid, and thick outer lithospheric shell—Venus decidedly is not. Neither is it under a plate-tectonics regime typified by Earth. But evidence for recent surface mobility (Byrne et al. 2021; Gascioli et al. 2025), together with considerations of interior heat production and flow and the high surface temperature, has led to the suggestion that the squishy-lid tectonic regime proposed for Archean Earth (Lourenço et al. 2020) may describe at least the Venus lowlands, too.

One of the biggest surprises of NASA's Magellan radar mission to Venus in the 1990s was the discovery of fewer than 1000 impact craters on the planet, seemingly distributed evenly across the surface (Phillips et al. 1991). The average model age for the planet's surface has been calculated at somewhere around 750 Myr (McKinnon et al. 1997), but there is substantial uncertainty in that estimate. Moreover, whether the spatial distribution of impact craters truly is random or merely seems so remains unknown (Phillips et al. 1992; Price et al. 1996). As a result, interpreting Venus' cratering record has led to two almost diametric hypotheses, one in which the planet was catastrophically resurfaced by volcanic activity over a relatively short period hundreds of millions ago (Schaber et al. 1992), and another in which Venus has been subject to steady-state volcanism throughout its preserved geological history (Phillips et al. 1992).

Efforts to decode Venus' geological history encompass, but extend beyond, whether it underwent catastrophic resurfacing or had a mobile-lid tectonic regime—although both topics tie into the broader question of the planet's potential past habitability. The elevated atmospheric D/H ratio has two leading interpretations: that either Venus lost substantial water in the form of steam from a primordial atmosphere degassed during the magma ocean phase (e.g., Gillmann et al. 2009; Hamano et al. 2013), or in fact *did* have oceans at some point, before losing them to evaporation and photodissociation having entered a runaway greenhouse state (Donahue et al. 1982; Way et al. 2016; Way and Del Genio 2020). Which of these scenarios is correct has profound implications not only for our understanding of Venus' geological and climate history, but of large rocky worlds generally... including our own.

If it is ultimately determined that Venus never cooled enough to condense its early, steam-rich atmosphere—say, because of its proximity to the Sun, compounded by thermal energy from impact bombardment, short-lived radiogenic elements, and so on—then we will have a ready test for the climate states of Earth-size worlds close to their stars: those worlds should be post-runaway greenhouses. If, on the other hand, we one day establish that Venus *was* truly Earth-like at some point in its history—a world with a clement climate and liquid water on the surface—then we will be faced with two consequences for interpreting rocky exoplanets. One: stellar distance may not matter as much as we once thought it did. And two: we will have to reckon with why Venus experienced such a drastic change to its climate.

Some hypotheses have been offered for the cause of Venus' potential past climate change. Perhaps multiple, huge volcanic eruptions—LIPs, to which Earth's Permian–Triassic mass extinction is attributed—were responsible, injecting so much $CO_2$ into the atmosphere in so short a time that the planet's presumed carbon–silicate cycle was overwhelmed (Way et al. 2022). Under this scenario, even with plate tectonics and subduction, Venus would have been unable to sequester enough $CO_2$ to prevent its buildup in the atmosphere, leading (perhaps geologically rapidly) to ever higher surface temperatures and the loss of its oceans. Eventually, without liquid water, subduction and thus plate tectonics would halt (Bolfan-Casanova 2007; Korenaga 2020) even as interior melting and volcanism persisted. It is not difficult to imagine a desiccating world with now-empty ocean basins yet high-standing terrain transitioning from the bimodal hypsometry of Earth to the unimodal signature of modern Venus as those basins filled with lava (Fig. 3).

There is, notably, an additional lesson for exoplanetary science here. As oceans evaporate within an increasingly warmer atmosphere, that water could extend to stratospheric altitudes above any hypothetical protective $O_3$ layer and become subject to photodissociation by incident UV stellar radiation (Ingersoll 1969; Kasting 1988). With H being lost to space, even for an Earth-size terrestrial planet, atmospheric $O_2$ abundance would rise for a time—supporting on the basis of disequilibrium chemistry an interpretation from afar of that world as habitable, and perhaps even inhabited, when in fact the opposite is true (Meadows 2017).

### 3.2 Mars

Perhaps the most obvious property of Mars relative to Earth and Venus is its substantially smaller size (0.53 $R_{\oplus}$) and lower mass (0.11 $M_{\oplus}$) (Table 1). Also notable is the planet's bimodal hypsometry (Fig. 3)—manifest as its north–south hemispherical dichotomy, where the northern third of the surface is topographically lower, has thinner crust, and is less cratered than the southern two-thirds. This dichotomy is attributed to degree-one mantle convection (e.g., Sleep 1994), to a positive feedback between crustal growth and mantle (Bonnet Gibet et al. 2022), or to a gigantic, ancient impact event (e.g., Andrews-Hanna et al. 2008), rather than to the presence of more buoyant, felsic rocks as is the case for Earth. Present surface conditions are very cold and extraordinarily dry, under a $CO_2$-dominated atmosphere with a surface pressure equivalent to an altitude of 45 km on Earth.

Contemporary activity includes marsquakes (as detected by NASA's InSight mission: Banerdt et al. 2020), great dust storms, and, of course, bolide impacts. But the majority of Mars' geological history was written a long time ago, a function of the planet's relatively diminutive size and the fact that its mantle appears to have degassed most of its (extricable) volatile inventory long before now (Grott et al. 2011). For all intents and purposes, the Red Planet is a quiescent, desert world today.

Nevertheless, the planet is replete with geological landforms we recognise from Earth (Fig. 8). There are vast shield volcanoes (Carr et al. 1977), the very biggest of which are concentrated into two enormous volcanic rises, Tharsis (the largest) and Elysium; most of these edifices have large summit caldera complexes. Basaltic lava flows are ubiquitous, and constitute much of the geologically younger units on the planet as well as probably the older, much more cratered rocks (e.g., Tanaka et al. 2014). Giant fluvial outflow systems course from equatorial latitudes to the northern lowlands, and were likely emplaced by catastrophic, rapid groundwater release (Baker 1979); smaller valley networks especially in the ancient southern uplands are interpreted to have been caused by surface runoff from rainfall. Tectonic landforms are widespread, the largest examples of which are in the southern hemisphere and are probably the result of global contraction from secular interior cooling (e.g., Nahm and Schultz 2011) (see Sect. 3.3). Extensional landforms are common, especially radial to the Tharsis rise (and are possibly underlain by dykes). Impact features abound and, at smaller spatial scales, sedimentary sequences corresponding to palaeolacustrine environments hosted within craters have been identified from both space and on the ground with robotic geologists (e.g., Mangold et al. 2021). Ice caps, chiefly of water in the north and carbon dioxide in the south, sit atop the poles.

The geological history of Mars is divided into three major eons: the Noachian (4.1–3.7 Ga), the Hesperian (3.7 Ga to approximately 3.0 Ga), and the Amazonian (around 3.0 Ga to today). That time before the Noachian is referred to as the pre-Noachian, although relatively little of that period remains, save for perhaps the planet's crustal dichotomy (Cassata et al. 2018) and hints of volcanic processes from Martian meteorites (e.g., McCubbin et al. 2016). Most of Mars' secondary crustal *volume* (if not the surface itself), probably dates to this time, too (Fig. 4).

The Noachian was characterised by major volcanic and tectonic activity, including the construction of the Tharsis and Elysium rises and their constituent volcanoes (Werner 2009), the development of the vast Valles Marineris canyon system along the planet's equator (e.g., Quantin et al. 2004; Brustel et al. 2017), and the onset of crustal shortening especially in the southern hemisphere. Mantle temperatures in the post-accretion period through some of the Noachian were sufficiently high to drive this activity, and certainly rifting, valley formation, and global contraction continued into the Hesperian. Even the Amazonian saw the emplacement of major plains units across the northern lowlands (Tanaka et al. 2014).

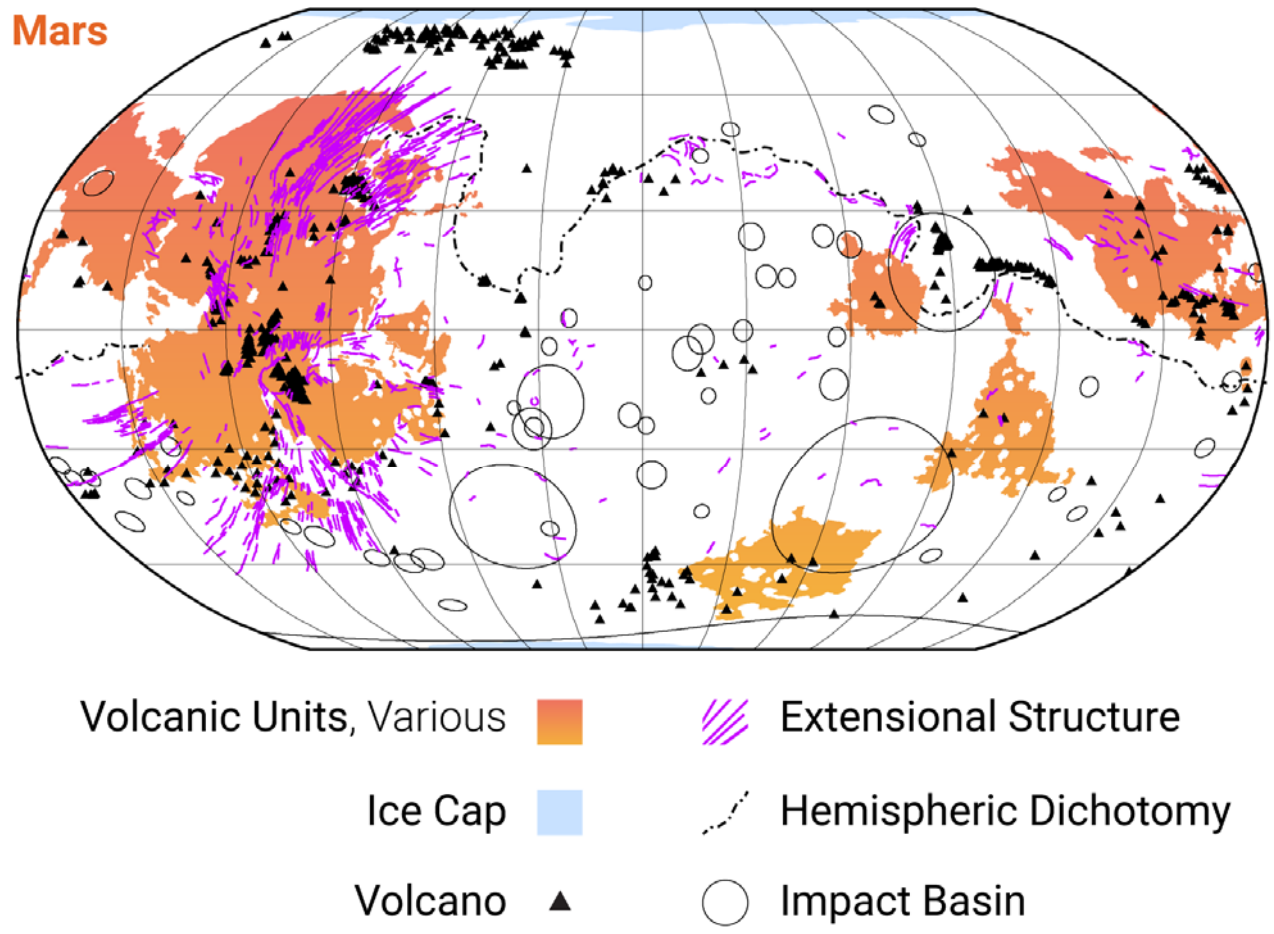

**Fig. 8** A global map of major geological features on Mars. Known volcanic units and mapped edifices are shown, although much of the remaining crust was probably originally volcanic. Extensional tectonic systems are shown, the majority of which are collocated with the Tharsis volcanic province in the planet's western hemisphere. Impact basins greater than 200 km in diameter are also included. There is no geomorphological evidence for Mars ever having had plate tectonics. Adapted from Byrne (2020) and references therein; the ice caps and crustal dichotomy boundary are from Tanaka et al. (2014)

But there is no evidence for major subduction or transform faults that characterise plate boundaries on Earth—telling us that Mars faced strict limits on its ability to engage in long-term volatile cycling necessary for a stable climate. Mars' smaller size (and thus greater ratio of surface to volume) led to relatively rapid cooling of the interior and the growth of a substantially thick lithosphere believed to have started to develop in the mid-to-late Noachian (Johnson and Phillips 2005). For most if not all of its geological history, Mars has been under a stagnant-lid regime.

The history of Mars' climate is far from fully understood (Kite and Conway 2024). The debate mostly relates to whether the planet had a long-term stable, temperate climate with surface liquid water for hundreds of millions of years, or was instead mostly an ice-house world with punctuated, short-term temperate periods and only spatially and temporally limited surface liquid water (cf. Wordsworth 2016). Recent three-dimensional general circulation modelling has shown that it is possible to have large lakes—or even a northern ocean, filled with water possibly erupted from the subsurface at low latitudes and carried northwards by the outflow channels—remain on the surface for extended periods (Kamada et al. 2020; Guzewich et al. 2021; Schmidt et al. 2022). Yet other such modelling supports an icy-highlands hypothesis (Palumbo et al. 2018), a generally cold Mars with only short, temperate intervals during which liquid water was stable.

Indeed, for many years it was not clear as to how Mars could have *ever* been warm enough to support liquid water on its surface since it receives only 43% of Earth's insolation today, and 4.2 Ga would have received even less—a Martian parallel to Earth's Faint Young

Sun Paradox. A variety of theories was proposed including high-level clouds, large $CO_2$ abundances, and the presence of other greenhouse gases—with a recent study suggested that large impactors early in Mars' history could have produced copious amounts of $H_2$, a potent greenhouse gas, in their thermal plumes (Haberle et al. 2019; Woo et al. 2019) in an otherwise $CO_2$-dominated, 1–2-bar atmosphere. It remains to be seen how long such warming conditions could last, given the lack of volatile cycling and Mars' modest atmospheric escape velocity for low-mass molecules such as $H_2$. In other words, the evidence for liquid water being stable on the Martian surface for at least *some* geological time and the planet's climate history remain to be reconciled.

### 3.3 Mercury and the Moon

Mercury and Earth's Moon are planetary bodies with heavily cratered and therefore very old surfaces, minimal atmospheres, and with most of their geological activity having taken place in the first billion years of Solar System history. The details of each world differ considerably—in no small part because the Moon coalesced from material ejected by the catastrophic collision between proto-Earth and Theia, and so has a genesis unlike any of the other rocky bodies we consider here (Canup and Asphaug 2001). Even so, it is because of these *general* similarities that we group them in this discussion. (It is also worth bearing in mind that, as the smallest rocky worlds in the inner Solar System, their similarly sized exoplanet cognates will continue to remain very difficult to detect for some time.)

Both of these worlds are relatively small (Fig. 2), with radii of 2440 km (0.38 $R_\oplus$) and 1740 km (0.27 $R_\oplus$) for Mercury and the Moon, respectively (Table 1), and both are low mass (0.06 $M_\oplus$) and (0.01 $M_\oplus$), respectively. Both have spin–orbit resonances, Mercury in a 3:2 ratio and the Moon in a 1:1 resonance that sees (thanks to libration, actually slightly more than) one side permanently facing Earth; we term that face the nearside). Both show long impact bombardment histories, and tectonic and volcanic signatures indicating that they have operated under a stagnant-lid regime as far back as their geological records extend (see Fig. 9 and Fig. 10). And both have extremely thin atmospheres—termed exospheres—that are entirely collisionless, unlike the bulk of Earth's substantial atmosphere. These exospheres consist of a sparse collection of atoms and molecules, chiefly derived from the surface by micrometeorite bombardment, that can readily escape into space.

The first solid surface to form on the Moon was that which crystallized from its magma ocean around 4.35 Ga (Borg and Carlson 2023), following the proto-Earth–Theia collision (Canup 2014). Whereas minerals such as olivine, orthopyroxene, and clinopyroxene sank through the magma ocean to settle and form the thick lunar mantle, anorthosite floated atop the remaining melt (e.g., Smith et al. 1970; Warren 1990). Today, this crust remains readily visible, comprising the heavily cratered and rugged lunar highlands (Fig. 4). A similar primary crust, this time enriched in carbon, may have formed on Mercury as low-density graphite crystallised and came to rest atop its final, global magma ocean (Vander Kaaden and McCubbin 2015). Evidence for this ancient flotation crust is supported by the presence of low-reflectance material on the surface (Peplowski et al. 2016). However, much of this primary graphite was probably destroyed by impacts or buried under volcanic deposits that formed the secondary, silicate crust.

Indeed, the volcanic history of Mercury proper began after the magma-ocean phase and featured voluminous effusive eruptions that quickly built up the planet's crust (Byrne et al. 2018a) (Fig. 4). Emplaced during ongoing impacts, those volcanic plains acquired a heavy cratering record and are termed the intercrater plains (Strom et al. 1975; Whitten et al. 2014). Effusive volcanism continued as impact bombardment waned, such that the last major plains units remained relatively unscarred; these comprise the overwhelming major of the planet's smooth plains, with some small amount made of impact melt and fluidized ejecta.

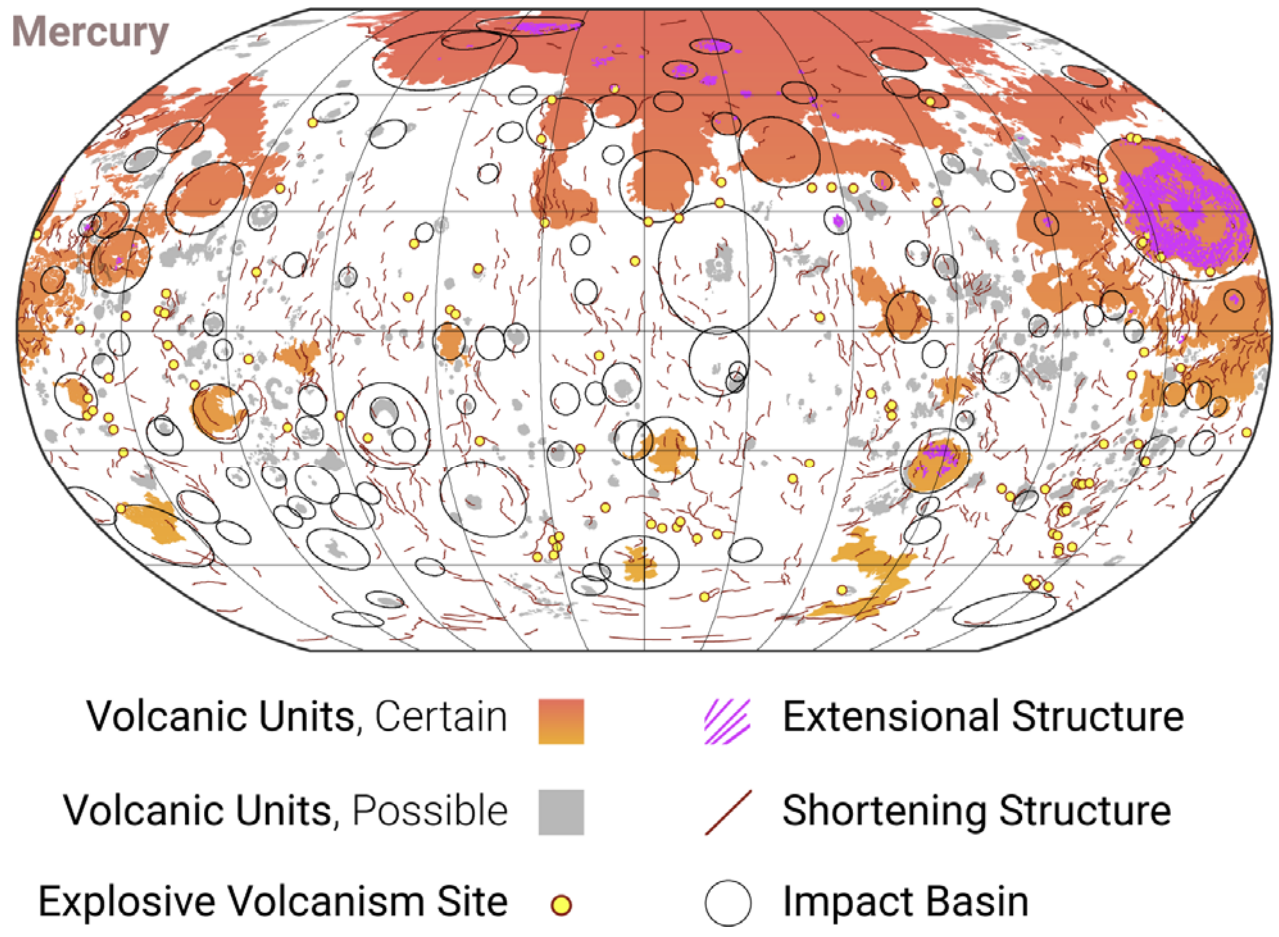

**Fig. 9** A global map of major geological features on Mercury. About a quarter of the planet surface comprises smooth plains, most of which by area and volume are known to be volcanic. Mercury has no resolvable plate boundaries nor constructional volcanic landforms, and virtually all extensional tectonic structures are situated within older volcanic plains deposits. Large thrust-fault-related structures are widespread, with a general difference in style between those in the smooth plains and in the older, cratered plains. Some explosive volcanic vents and/or deposits are recognised, most of which are collocated with sites of major crustal weakness (e.g., faults). Impact basins greater than 200 km in diameter are also shown. Adapted from Byrne (2020) and references therein, with the shortening structures from Byrne et al. (2014)

The major crust-building phase ended around 3.5 Ga, which is when the planet entered a state of global contraction from secular interior cooling (Byrne et al. 2016) (Fig. 4). This phenomenon led to the lithosphere entering a state of horizontal compression that first reduced and then shut off major volcanic activity, accounting both for why the last voluminous volcanic plains units are situated within sites of crustal weakness and reduced overburden—chiefly, impact basins—and why there are no large shield volcanoes or caldera complexes as are observed on Earth, Venus, and Mars (Byrne 2020). Explosive volcanism, on the other hand, outlasted effusive activity and endured at least until 1 Ga (Thomas et al. 2014). As for effusively emplaced plains, most explosive volcanism is spatially associated with areas of crustal weakness such as impact features or major tectonic structures.

The tectonic history of Mercury is characterised by this global contraction, which produced a worldwide network of large thrust-faut-related landforms (Byrne et al. 2018b). The total radial contraction of Mercury is at least several kilometres (Loveless and Klimczak 2025). Much of this deformation began around 3.5 Ga (Byrne et al. 2018b), but small extensional structures, only a few dozen meters tall and several kilometres long, atop larger thrust faults formed geologically recently and remain visible even in the face of low-level but constant impact bombardment, indicating that Mercury continues to contract into the present. Extensional deformation is otherwise almost entirely restricted to volcanically flooded impact basins (Fig. 9). Whether the relatively thin mantle is still convecting remains unclear (e.g., Jain and Solomatov 2024).

Secondary volcanism on the Moon started with the formation of the so-called Mg-suite, the preponderance of which has radiometrically deduced absolute ages of ca. 4.3–4.36 Gyr (Borg and Carlson 2023), almost certainly at or after the end of magma ocean solidification

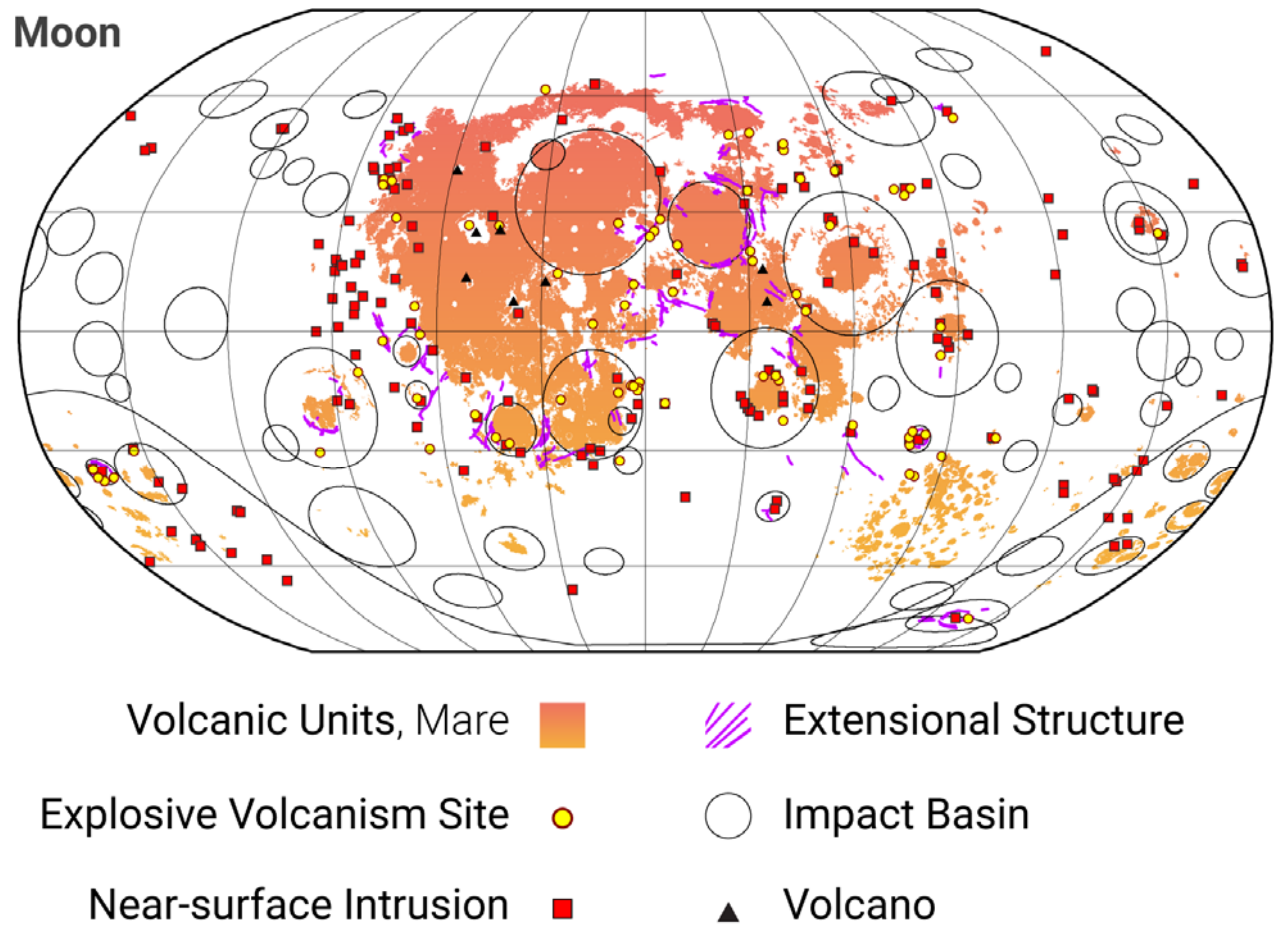

**Fig. 10** A global map of major geological features on the Moon. The anorthositic flotation crust makes up most of the lunar surface, with geologically younger volcanic plains (the "maria") mainly though not exclusively situated within recognised, antecedent impact basins. Sites of explosive and near-surface volcanic activity are shown, as are impact basins 200 km in diameter and above. The Moon has a handful of broad but extremely topographically subtle shield volcanoes, and some extensional structures (most of which are associated with impact features). The outer, brittle lithosphere of the Moon appears to always have acted as a single shell. Adapted from Byrne (2020) and references therein

and possibly as a consequence of large-scale overturn of that crystallised and gravitationally unstable magma ocean. A second phase of intense volcanic activity on the Moon occurred primarily between 3.8 and 3.2 Ga, resulting in the vast basaltic plains known as lunar maria, which were formed by effusive lava flows filling vast, antecedent impact basins (Fig. 10).

Eruptive activity continued, albeit at a much smaller rate, especially in the so-called Procellarum KREEP Terrain (PKT, for potassium-, rare-Earth-element, and phosphorus-enriched radiogenic elements) on the western side of the Moon's nearside. On the basis of samples returned to Earth, (relatively) young basaltic ages up to 2 Gyr have been identified at the Chang'e 5 landing site in Procellarum (Li et al. 2021), and model ages of between 3 and 1 Gyr have been suggested from crater statistics of nearside maria (Hiesinger et al. 2003). The formation of the PKT is a point of some controversy, but the observed strong abundances of radioactive elements in this region is likely genetically linked to its history of relatively more recent volcanism (Morota et al. 2011). Even so, other regions on the farside, too, show signs of some volcanic activity. For instance, the Chang'e-6 mission returned samples of mare basalts from the north-eastern South Pole–Aitken basin on the Moon's farside dated to 2.8 Gyr (Zhang et al. 2025).

Similarly to Mercury, tectonic activity on the Moon has been principally driven by the cooling and contraction of its interior, leading to the formation of thrust-fault-related structures across the lunar surface, albeit at magnitudes far below those on the innermost planet (Watters et al. 2019). Some additional crustal shortening is evident in the maria, in part the result of isostatic adjustments of mantle uplifted by early, giant impacts (Byrne et al. 2015). Extensional deformation is very modest on the Moon, generally spatially associated with impact features (and probably genetically related to their volcanic infills: Klimczak 2014).

Mercury and the Moon represent end members among terrestrial bodies in terms of their interior structure (Fig. 2 and Tables 1 and 2): the Moon has a small, iron-rich core ($R_{\mathrm{core}}$ $\sim$0.2 $R_{\mathrm{body}}$), whereas Mercury has an outsize core ($\sim$0.8 $R_{\mathrm{body}}$) that also generates a mag-netic field in the present. The Moon's relatively low mass fraction of iron may be the result of its coalescence in the aftermath of a giant impact (e.g., Canup and Asphaug 2001), although early in lunar history an internally generated dynamo was sufficiently strong to induce remanent magnetization in mare basalts, and may have operated for a few billion years (Wieczorek et al. 2023). In contrast to Mercury, the Moon has no modern magnetic field.

The formation of Mercury's large core is less well understood, with various mechanisms having been suggested such as the separation of metals and silicates in the solar nebula (Lewis 1972; McDonough and Yoshizaki 2021), the loss of a large fraction of its rocky mantle after its formation by evaporation followed by solar wind (Fegley and Cameron 1987), or by mantle stripping by one or several giant impacts (Benz et al. 1988; Chau et al. 2018).

Most of these scenarios tend to result in a volatile-poor interior for Mercury. However, measurements of elevated potassium–thorium and potassium–uranium ratios suggest that Mercury, despite its proximity to the Sun, is somewhat rich in volatile elements (Peplowski et al. 2011). Furthermore, geochemical mission data imply that the planet formed under very reduced conditions (e.g., Nittler and Weider 2019). Among the terrestrial worlds, the planet's oxygen fugacity is the lowest and its sulphur content in the mantle the highest. This unique chemical environment influences, in turn, the metal–silicate partitioning of typical lithophilic elements, particularly thorium and uranium. The result may be a greater core enrichment of U and Th (McCubbin et al. 2012)—which could imply that Mercury is more depleted in volatile elements than the current K/U and K/Th surface ratios would suggest. Such core enrichment, if true, also has implications for the evolution of the magnetic field.

Specifically, and in addition to its modern intrinsic magnetic field, the planet also had a dynamo prior to about 3.7 Ga, as shown by measured remnant magnetisation of basaltic crust at high northern latitudes (Johnson et al. 2015). It is currently unknown whether the same dynamo today has been active all this time, or whether the field was revived later in Mercury's evolution, for example by the growth of the solid inner core—which has been detected on the basis of libration and gravity measurements (Genova et al. 2019).

### 3.4 Io

We include a discussion of Io because, although it is not situated in the inner Solar System, it is a rocky body and boasts abundant volcanic and tectonic landforms. Moreover, it is the most geologically active body yet recognised in the outer Solar System, and is the only giant planet satellite not covered by a layer of water ice. By these measures, it warrants discussion here—especially given the relevance of its tidal heating (see below) to understanding rocky exoplanets close to their host stars and subject to strong tidal forcing (Sect. 4.2.2).

The innermost of the four Galilean moons of Jupiter, and with a radius of approximately 1800 km (0.29 $R_{\oplus}$), Io is slightly larger than the (our) Moon (Fig. 2). Io has an iron-rich core and a silicate mantle, with a metallic core of radius somewhere between 650 and 950 km (Sohl et al. 2002). Unlike the Moon, however, the surface of Io is crater free, with lava flows that cover virtually the entire surface (McEwen et al. 1998; Williams et al. 2011) and towering mountain blocks (Schenk et al. 2001) (Fig. 11). Chief among the satellite's most notable volcanotectonic landforms are paterae, broad depressions that are commonly filled with lavas and are widespread across the planet (Radebaugh et al. 2001).

Most of the surface is basaltic, but some lava flows are sulphur-rich and widespread sulphur-dioxide frost (condensed from erupted $SO_2$ particles) has also been observed (e.g., McEwen et al. 1988). Eruption temperatures of up to 1600 K have been measured, which can be explained by the effusion of high-temperature, ultramafic silicate lavas (Keszthelyi et al. 2007).

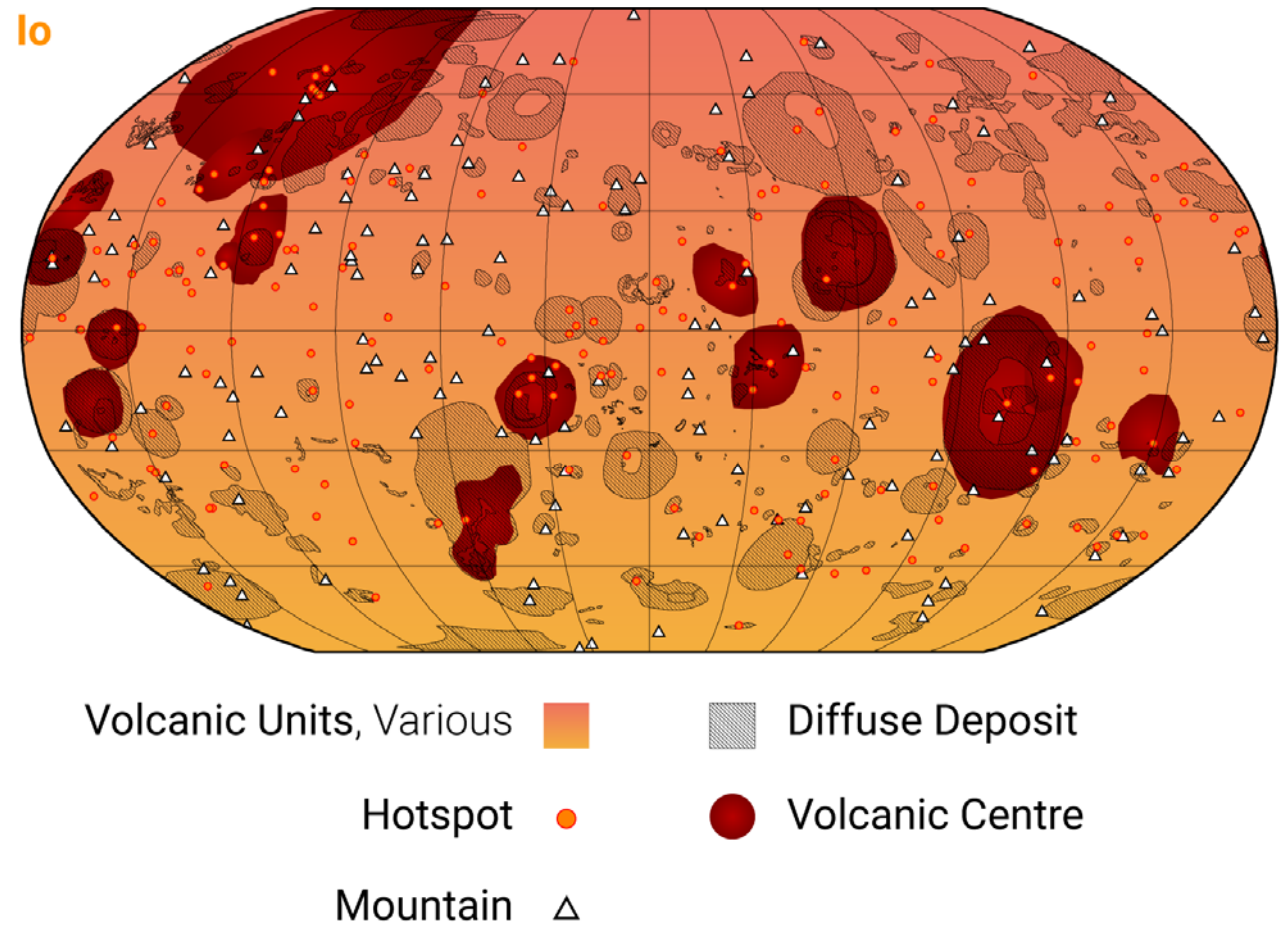

**Fig. 11** A global map of major geological features on Jupiter's volcanic moon Io. Virtually the entire surface is volcanic, although large mountain blocks are distributed across the body. Sites of effusive and explosive volcanic activity, Io's "paterae", are also shown—some of which are clustered into centres. Diffuse pyroclastic deposits are marked with a hachured pattern. Io has no confirmed impact features and no tectonic patterns indicative of a fragmented brittle lithosphere. Adapted from Williams et al. (2011) and references therein

Indeed, Io is the most volcanic active body in the Solar System, emitting between 1.5 and 4 $W/m^2$ of infrared radiation from the surface (e.g., Moore et al. 2007)—corresponding to a surface heat flux some 15 to 40 times greater than that of Earth. To explain this high heat flow, more energy is required than that released during planetary formation and silicate–iron differentiation (i.e., primordial energy) and through the decay of radioactive elements, as is the case with Earth and the other inner Solar System worlds. Instead, the internal heat energy driving Io's prodigious volcanism is principally supplied by tidal interactions with Jupiter and its neighbouring moons: Io experiences periodic, gravitational tides from Jupiter that generate strong internal friction because of its close distance to Jupiter and its eccentric orbit (Peale et al. 1979). The so-called Laplace resonance of the Galilean moons Io, Europa, and Ganymede (in a 4:2:1 ratio of orbital periods) sustains the eccentricity of Io's orbit and is essential for its strong tidal dissipation.

Io's resurfacing rate has been estimated to be $\sim$1.5 cm/yr (Kirchoff and McKinnon 2009), leading to subsidence of the crust under continually emplaced lavas (Schenk and Bulmer 1998). This subsidence, in turn, is predicted to give rise to a horizontally compressive stress state at the base of the Ionian lithosphere (Bland and McKinnon 2016), forming deep-rooted thrust faults along which uplifted crustal blocks rise to form the moon's gigantic mountains (Schenk and Bulmer 1998).

The observed thermal emission is not evenly distributed across the moon's surface. Active and recent paterae show higher emission and, although volcanism is distributed globally, it seems to be concentrated at lower latitudes (e.g., Kirchoff et al. 2011; Hamilton et al. 2013). Knowledge of the distribution of volcanic activity is important to understand where in the satellite tidal dissipation takes place. Deep tidal heating is predicted to produce greater heat flux at the poles, whereas shallow heating (e.g., in an asthenosphere) is thought to lead to greater heat flux near the equator (Segatz et al. 1988). Recent geophysical measurements from NASA's Juno mission indicate that voluminous melting at shallow depths within Io is unlikely, and that the mantle is likely mostly solid (Park et al. 2025). This inference is consistent with Juno observations of more volcanic activity at the poles than previously recognised (Pettine et al. 2024).

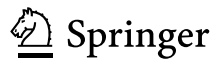

Even so, Io's internal structure as a consequence of tidal dissipation has likely evolved over time. The interplay between tidal dissipation and orbital resonance with Europa and Ganymede suggests that Io's volcanic activity and internal structure may vary over several-hundred-million-year timescales, with varying degrees of mantle melting and chemical layering (Hussmann and Spohn 2004). This assumption is supported by the current migration of Io towards Jupiter (Lainey et al. 2009) as a result of tidal heating removing energy from the moon's orbit. Here, the lesson is that tidally induced geological activity is subject to temporal changes in intensity, in a manner similar to (if not quite the same as) the long-term reduction in activity normally associated with worlds powered chiefly from within.

## 4 Looking Beyond the Solar System

### 4.1 Lessons Learned and Questions Arising from the Solar System

The rocky worlds of the Solar System tell us that, despite the fundamental similarities in physical make-up of terrestrial planetary bodies, the *details* of each vary substantially (Table 2). Earth has remained habitable (insofar as liquid water has been stable at its surface) for virtually all of its geological history starting from shortly after the Moon-forming impact—even though its geodynamical state, surface composition, atmospheric make-up and volume, and even its physical appearance have changed considerably since formation. Would we recognise Hadean Earth as our own world if we saw it from a neighbouring planetary system? Is plate tectonics a prerequisite for long-term habitability? Did the impact with Theia set Earth on a path that would otherwise have left it inhospitable to life?

Venus, too, is an enigma. The second-largest rocky body in the Solar System may once have had a clement climate and perhaps been like Proterozoic or even Phanerozoic Earth, in which case it, too, is an example of how the face of a rocky world can change dramatically over geological time. Alternatively, Venus' fate may have been set from the beginning, its distance from the Sun meaning that it could never be temperate. If so, then we might assume that most terrestrial exoplanets with atmospheres within the "Venus zone" (Kane et al. 2014) are, in fact, in post-runaway greenhouse states like Venus. It is for this reason that establishing whether, why, and when Venus diverged from the path taken by Earth is so important (Kane and Byrne 2024): what does Venus tell us about the rules that govern Earth-size worlds in general?

Whereas we have yet to unravel Venus, we know definitively that Mars has undergone major climate change through Solar System history. Where once were lakes, rivers, and an active hydrological cycle there is now a cold, dry world. The key to understanding Mars lies in its thermal evolution and eventual loss of an internally generated magnetic field that was unable to prevent its atmosphere from being stripped by the solar wind (Jakosky et al. 2018). Yet there is an equally important element here: the ability of the mantle to *replenish* that lost atmosphere by volcanic degassing. With a mantle and crust together only 15% the size of that of Earth (Byrne 2020), Mars functionally exhausted its degassable volatile budget long ago. Thus, its climate history reflects both its stellar environment and its own geological and physical properties.

Mercury and the Moon share broadly similar geological histories despite differences in size, mass, and distance from the Sun. Both worlds cooled early on; their surfaces have remained largely unchanged since the first billion years after formation. Volcanism was prodigious on Mercury, before waning as interior cooling and global contraction took hold. The

**Table 2** Key geological properties of the inner Solar System worlds. All data are from publicly available sources except for interior structure information, which is taken from Byrne (2020) and references therein

| | Mercury | Venus | Earth | Moon | Mars |
|---|---|---|---|---|---|
| Magnetic Field? | Intrinsic | Induced | Intrinsic | Remanent | Remanent, Induced |
| Atmosphere? | No (Exosphere) | Yes | Yes | No (Exosphere) | Yes |
| Surface Pressure (Bar) | $10^{-12}$ | 92 | 1 | $3 \times 10^{-15}$ | $6 \times 10^{-3}$ |
| Atmosphere/ Exo-sphereComposition | • H, $H_2$ (Trace)<br>• O, $O_2$ (Trace)<br>• Na (Trace)<br>• Ca (Trace) | • $CO_2$ (96.5%)<br>• $N_2$ (3.5%)<br>• $SO_2$ (Trace)<br>• Ar (Trace) | • $N_2$ (78.1%)<br>• $O_2$ (21%)<br>• Ar (0.9%)<br>• $CO_2$ (Trace) | • Ar (Trace)<br>• He (Trace)<br>• Ne (Trace)<br>• Na (Trace) | • $CO_2$ (95%)<br>• $N_2$ (2.8%)<br>• Ar (2%)<br>• $O_2$ (Trace) |
| Temperature Range (°C) | −173 to 427 | 450 to 475 | −89 to 57 | −133 to 121 | −153 to 20 |
| Mean Temperature (°C) | 167 | 464 | 15 | −21 | −65 |
| Liquid Surface Water? | No | No | Yes | No | No |
| Water Ice? | Yes (In Polar Craters) | No | Yes | Yes (In Polar Craters) | Yes |
| Cessation of major plains volcanism (Ma) | 3500 | ∼750 | N/A; ongoing | 2500 | 1600 |
| Radius reduction from interior cooling (km) | 4.7–7.1 | ? | ? | >0.07 | 0.2–3.8 |
| Global, late-stage plains volcanism? | No | Yes | Yes | No | Yes |
| Plains volcanism within impact basins? | Yes | No | No | Yes | No |
| Large shield volcanoes? | No | Yes | Yes | No | Yes |
| Major extensional structures/rifts? | No | Yes | Yes | No | Yes |

Moon's volcanic record is primarily constrained to ancient impact structures. Neither world possesses a large silicate portion (indeed, their mantle volumes are within about 20% of each other: Byrne 2020), consistent with early-onset cooling and an inability to replenish an atmosphere. The innermost planet, especially, remains to be fully understood given its surprisingly high core mass fraction. The formation and subsequent evolution of rocky planets with outsize cores is important to understand, as the discovery of so-called super-Mercuries (Sect. 4.2.6) in other planetary systems (e.g., Lam et al. 2021) suggests that they could be more common than we might have expected.

Io merits a mention here, too, since it is a world for which we would have no expectation of major geological activity were it not for the remarkable degree to which it is tidally heated. This process may play a key role in the thermal evolution of rocky exoplanets that are close to their host star and subject to strong, sustained tidal forcing via mean motion resonances. Enhanced, sustained solid-body tidal dissipation could help keep the interiors of even modestly sized worlds warmer than they might otherwise be (e.g., Makarov et al. 2018; Revol et al. 2023), offsetting the slow but irrevocable cooling that defines the pathways taken by Mars, Mercury, and the Moon. Io is therefore an ideal laboratory for improving our understanding of this likely common process.

## 4.2 What Are We Missing?

Although the worlds in the Solar System may be diverse, we have come to realise that they represent but a portion of the broader tapestry of possible rocky planet outcomes (Fig. 1). This observed diversity of exoplanets is equal parts tantalising and frustrating. On the one hand, we now know that other possible, fantastic places do exist. On the other, these worlds for so far outside the reach of conventional spacecraft that we must settle with the few bulk planetary properties we can realistically measure, inferences from which have too-often-degenerate explanations (Sect. 1.2). It is tempting to contextualise newfound planets against the known yardsticks of our Solar System neighbours, evinced in nomenclature by terms such as "super-Earths" and "sub-Neptunes", even if those alien worlds are utterly unlike Earth and Neptune. Thus the true extent of the (dis)similarities between what we know and what we are discovering, although highly motivating, remains far from established.

This section briefly reviews some prominent, overlapping subsets of the exoplanet population for which we are missing an obvious Solar System analogue. Although our discussion here is restricted to terrestrial (and thus modestly sized) worlds, giant exoplanets also appear to span a wider parameter space than that of Jupiter, Saturn, and Neptune and Uranus (e.g., Fig. 1). We also caution that we should expect substantial variation between individual planets within the same "class", just as there is between Earth and Venus, because of diverging starting and/or evolving conditions that, absent direct observations or measurements, are almost impossible to predict accurately (Jakosky and Byrne 2025). Even so, we can make broad inferences regarding how these worlds may look, and operate, based on our knowledge of the inner Solar System's planetary inventory.

### 4.2.1 Super-Earths

In practice, the term super-Earth usually means a planet of ∼1–1.6 Earth radii ($R_\oplus$) (Fig. 1), larger than Earth but below the "radius valley"—a term used to describe the empirically observed deficit in exoplanets with radii at roughly 1.5–2 $R_\oplus$, depending on orbital period and host star's spectral type (Fulton et al. 2017; Gaidos et al. 2024). Most (but not all) planets with observed radii below the valley have bulk densities consistent with a rock–iron composition, whereas most (but not all) planets with radii above the valley have low bulk densities that require a non-negligible mass fraction of lighter gases or ices (Rogers 2015; Baumeister et al. 2025, this collection). The radius valley is generally interpreted as marking a divide between rocky worlds and sub-Neptunes, but like many others this term does not have a robust and consistent definition.

Nevertheless, the general view is that, like our own inner Solar System worlds, super-Earths are principally built from rock and metal, perhaps with a secondary (volcanically outgassed) atmosphere that comprises a relatively small ($\ll$1%) proportion of the planet mass. Some super-Earths are almost certainly bare with no atmosphere (e.g., LHS 3844 b: Lyu et al. 2024) and thus represent truly massive terrestrial worlds the likes of which we do not have in the Solar System. Yet some super-Earths have a low bulk density that indicates a substantial non-rocky component, such as a thick but chemically light atmosphere, and so the silicate and metal portion may be more modest and closer to Earth or Venus. The exoplanet L 98-59 d (Demangeon et al. 2021) is one such example, with one of the best-known super-Earths, 55 Cancri e, another (Demory et al. 2011).

It may be that some super-Earth-size exoplanets are the remnant rocky cores of formerly puffier planets that have lost their primordial hydrogen atmospheres. This line of progeny offers one account for the radius valley: planets formed of a size 1.5–2 $R_\oplus$ could then lose much of their primordial hydrogen envelope to stellar wind stripping, such that they experience a reduction in radius over time and so move to below the valley (Rogers and Owen 2021).

One paradigm-shifting corollary of this hypothesis is the prospect that some super-Earths today formed with substantial hydrogen atmospheres. It is interesting to consider whether observational tests could distinguish a planet that originated in this way from one "born rocky"—say, for example, if the lost hydrogen has chemically imprinted upon the planet (Kite and Schaefer 2021). A second such corollary is that the loss of a substantive atmosphere to eventually expose the silicate-and-metal world beneath would essentially lead to the formation of a rocky planet potentially several *billion* years after the planetary system has established (Nicholls et al. 2025). A terrestrial world thusly formed would presumably lack the impact bombardment history of the Solar System's inner bodies but might otherwise be geologically active. Indeed, the thermally blanketing effect of the (former) $H_2$ envelope could help offset interior cooling and global contraction for far longer than would be the case for a more modest atmosphere, even if some of this hydrogen were ingassed (Nicholls et al. 2025).

Super-Earths have naturally served as a starting point for extrapolating Earth-based geological processes, in that they allow us to ask questions of the form: "how does a given geophysical process scale with planetary mass?" Two decades' work has set a sizeable—if not unanimous—theoretical foundation of studies on the possible mantle dynamics (e.g., Stamenković et al. 2012; Tackley et al. 2013; van den Berg et al. 2019), tectonic regimes (e.g., Korenaga 2010; Foley et al. 2012), and interior oxidation states (Elkins-Tanton and Seager 2008; Lichtenberg 2021) of super-Earths. With so little information to go on, and given that the category "super-Earth" likely encompasses multiple different types of world in the first place (Lichtenberg and Miguel 2025), there is considerable degeneracy in our ability to truly understand these worlds. Nevertheless, if the rather fantastical idea of a super-Earth orbiting the Sun in the far outer Solar System (Sheppard and Trujillo 2016) is true, we might someday rejoice in exploring a (relatively) nearby example.

### 4.2.2 Rocky Planets Tidally Locked to Their Stars

Tidally locked moons are common in the Solar System, but planets tidally locked to the Sun are not. Perhaps the best known such planetary example in our planetary system is Mercury, which rotates thrice for every two orbits around the star, in contrast to the canonical 1:1 spin–orbit resonance that is typified by the Moon (Sect. 3.3). But exoplanets close to their stars readily become tidally locked or near-locked, with the dayside (semi-)permanently facing their sun (e.g., Revol et al. 2024; Lyu et al. 2024). This configuration is especially relevant for planets orbiting M-dwarf stars, where tidal-locking-prone orbital periods fall within the classical circumstellar habitable zone.

Modelling studies of tidally locked rocky worlds (e.g., Pierrehumbert 2011; Turbet et al. 2016; Del Genio et al. 2019; Leconte 2018; Way 2025) suggest, as a function of atmospheric, ocean, and/or geodynamic properties, that substantially different climate and geological outcomes to Earth and Venus are possible—such as hemisphere-scale differences in volatile deposition (Pierrehumbert 2011) or in tectonic and volcanic activity (Meier et al. 2021). The bulk physical properties of stellar-tidally locked worlds might therefore resemble those with which we are familiar, but their surfaces and climates could be vastly different.

#### 4.2.3 Lava Worlds

Some potentially rocky exoplanets orbit close enough to their stars to reach rock-melting or even rock-*evaporating* temperatures at their dayside surfaces (Chao et al. 2021). These worlds are therefore believed to host long-lived magma oceans (Léger et al. 2011), which could also be sustained on geological timescales by tidal heating (Nicholls et al. 2025). Pending any detection of telltale identifiers of an atmosphere (e.g., Zilinskas et al. 2022; Teske et al. 2025), the calculated radiative equilibrium temperature is the criterion used to infer the presence of a surface magma ocean on exoplanets. Temperatures this extreme do not exist on any surface in the contemporary Solar System, save for transiently as lava lakes on Earth and Io (Lopes et al. 2018).

Meanwhile, on young planets, magma oceans arising from strong heating (such as accretion, impact bombardment, short-lived radiogenics, and differentiation) are thought to set these worlds' long-term conditions, in part because molten states permit rapid thermal and chemical equilibrium (Elkins-Tanton 2012; Hamano et al. 2013; Deng et al. 2020; Lichtenberg et al. 2023). Lava-world exoplanets resemble these pivotal, primordial magma oceans but for three notable ways:

1) If a planet receives enough instellation to render its surface molten by virtue of its close proximity to its host star, then that world is likely tidally locked;

2) Lava planets are probably heated more strongly from above than from within, affecting the maximum depth and patterns of convection within the molten layer and, possibly, the solid-but-plastic silicate layer below; and

3) Whereas primordial magma oceans endured in the Solar System for fewer than perhaps 100 million years (the exact timings of which are still uncertain), magma oceans on lava worlds may persist for many times that. Over this time, silicate vaporisation and transport, degassing, and atmosphere–magma redox exchange could drive chemical differentiation under some conditions (Kite et al. 2016; Curry et al. 2024; Boukare et al. 2025) that are not replicated in the Solar System.

Because they are so hot (Greene et al. 2023), lava worlds are relatively amenable to secondary eclipse observations of their thermal emission, which could in principle identify both volatile- and silicate vapour species in their atmospheres (Zilinskas et al. 2022; Piette et al. 2023). These bodies are intriguing because they may offer unprecedented insight, through thermal emission spectroscopy, into bulk rock compositions (subject to the third point, above) not otherwise available for rocky exoplanets (Seidler et al. 2024).

#### 4.2.4 Long-Lived Internally Heated Planets

It is possible for rocky exoplanets to also have much stronger and longer-lived *internal* heat sources than those present in Solar System planets. We distinguish these planets from lava worlds in that the surface need not be molten. Contending long-lived heat sources include: tidal dissipation (e.g., Jackson et al. 2008; Hay and Matsuyama 2019); high abundances of radiogenic nucleotides (Nimmo et al. 2020) as a consequence of the composition of the molecular cloud from which the planet (and system) formed; and magnetic induction heating (i.e., Ohmic dissipation: Bromley and Kenyon 2019; Kislyakova and Noack 2020).

Because they have more energy to lose to space, we can expect such bodies to remain geodynamically active for longer, all other things being equal: their mantles persist in convecting and partially melting for more time than, say, would any of the inner Solar System worlds (Unterborn et al. 2022). By this measure, even smaller worlds the size of Mercury can be expected to display surfaces of the complexity of Mars or even of Venus or Earth (Byrne 2020), resisting the onset of global contraction and preserving for longer the ability to replenish atmospheric loss and thus retain resolvable atmospheres for multiple billion years (assuming sufficient volatile inventories) (Jakosky and Byrne 2025).

#### 4.2.5 Very Young and Very Old Planets

We assume that the age of a planet is functionally the same as that of its host star, given that both form quickly from the same molecular cloud, and we estimate the age of that star through a variety of means, including luminosity and comparison with other stars in the same cluster. Although the errors associated with stellar age estimates are often large, even loosely bound such ages offer a new axis of parameter space. The growing population of planets less than 1 Gyr old is an opportunity to see how very young planets appear. On the basis of what we see in the Solar System, we should be careful to not assume planets of a given type evolve on a common trajectory (cf. Jakosky and Byrne 2025), but young worlds nonetheless present a way to test hypotheses and predictions of early planet evolution involving magma ocean solidification, primordial atmosphere retention and replacement, formation of secondary crust, and so on. Equally, there are other rocky planets likely several billion years *older* than Earth (e.g., Burgasser and Mamajek 2017). Finding signs of volcanism, or even the presence of an atmosphere at all, on such worlds would empirically delineate their geodynamical and climate histories, especially regarding whether it is the fate of all large, rocky planets to acquire a post-runaway greenhouse state (Boukrouche et al. 2021).

#### 4.2.6 Super-Mercuries

Several near-Earth-radius exoplanets have measured masses that, if correct, mean they are far denser than Earth. The most plausible explanation is that these planets are strongly enriched in iron—which presumably means an oversize core, akin to Mercury (Valencia et al. 2006; Wagner et al. 2012). Although it remains unclear how common these planets truly are, seeing as mass and radius measurements are still being revised, super-Mercuries would represent an intriguing outcome of planet formation. Their occurrence would help place bounds on models for Mercury's own formation, including whether its unusually large core mass fraction results from accretion in a highly reduced inner protoplanetary disk or from impact stripping early in its life (e.g., Wurm et al. 2013; Dou et al. 2024). Moreover, since Mercury's geological history has been dominated by global contraction from secular interior cooling (Byrne et al. 2018b), so too might super-Mercuries—leading to relatively short-lived volcanic (if not *magmatic*) activity (Byrne et al. 2016) and so limited ability for atmospheric replenishment by volcanic degassing (Noack et al. 2014).

#### 4.2.7 Super-Ganymedes

A population of cool planets shows radii similar to or larger than Earth, but of lower mass (Fig. 1). One reasonable interpretation is that these bodies are covered by thick water/ice layers in some ways similar to outer Solar System moons. These low-density worlds challenge the view that only rocky worlds can be potentially habitable—an issue we face in our own planetary system in terms of where the "habitable zone" is, given the preponderance of known or suspected subsurface oceans in the outer Solar System—and that smaller worlds are always rocky. These putative Super-Ganymedes invite the application of geodynamic models to non-silicate materials—that is, to ice (Lebec et al. 2023), and vice versa, of glacier dynamical models for Earth to planetary ice shells (Law 2025). In doing so, we might better understand how worlds with thick $H_2O$ layers (be they liquid and/or ice) such as our own Ganymede (Vance et al. 2018) internally transport heat and nutrients.

#### 4.2.8 Non-'chondritic' Rocky Planets

Most rocky bodies in the Solar System have a bulk refractory and moderately volatile element composition roughly similar to chondritic meteorites, thought to represent the building blocks of the planets (e.g., Palme et al. 2014). Yet, given the known variability of elemental abundances across stars, the "chondrite" counterparts of another planetary system will probably not exactly match our own (Wang et al. 2018), even if they are broadly similar. And so it may be that planets in other systems could have bulk compositions less consistent with a Solar System chondrite (Guimond et al. 2024).

To wit, processes in the protoplanetary disk can modulate a planet's rock-forming element abundances away from that of its host star. Potentially iron-enriched super-Mercuries are one example, but there are others: close-in planets may form enriched in calcium and aluminium, which would result in a measurably lower bulk density (Dorn et al. 2019; Hatalova et al. 2025). An early study proposed a carbon-rich composition to explain the low bulk density of 55 Cancri e (Madhusudhan et al. 2012), although cosmochemical models later showed this scenario to require stars with C/O ratios of $\sim>0.8$ (Larimer 1975), which we now believe are rare (Teske et al. 2013). Nevertheless, even every small changes in volatile composition—say, water abundance (Guimond et al. 2026, this collection)—could lead to vastly dissimilar bulk, geodynamical, geological, and climate outcomes.

### 4.3 Concluding Thoughts

Perhaps the fundamental lesson we should take from all of the observations we discuss here is that planetary bodies *change*. We summarise those changes, as they are currently understood, for the inner Solar System worlds in Fig. 12. Even with several decades of planetary exploration, note how many uncertainties feature in that figure—especially for Venus.

Nevertheless, we might draw some basic rules of thumb from the inner Solar System: that relatively larger planets several billion years old are more likely to possess moderate or even substantial atmospheres, recent or ongoing volcanism (enough even to offset atmosphere that is lost to space or to weathering reactions with rock), and maybe even a magnetic field and surface conditions amenable to liquid water. Meanwhile, smaller worlds are fated to cool, contract, and so lose most of their geological activity faster.

Yet tidal (or induction) heating complicates even this elementary guidance. So, too, does the stochastic nature of planetary evolution, evinced for example by the fact that giant collisions seem to have influenced the inner Solar System worlds to varying degrees. In other systems, different orbital distances and arrangements will effect different outcomes in impact bombardment, orbital obliquity (cf. Laskar and Robutel 1993), tidal heating, geology (e.g., Byrne et al. 2021), and even available feedstock during planetary formation. On that last point, *when* and *where* in the galaxy a planet forms may control its starting composition and elemental abundances, although to what extent remains a wide-open question. All of this is to say that, with minimal information about a *specific* world, we are limited in what we can predict—and how confidently we can predict it (Jakosky and Byrne 2025).

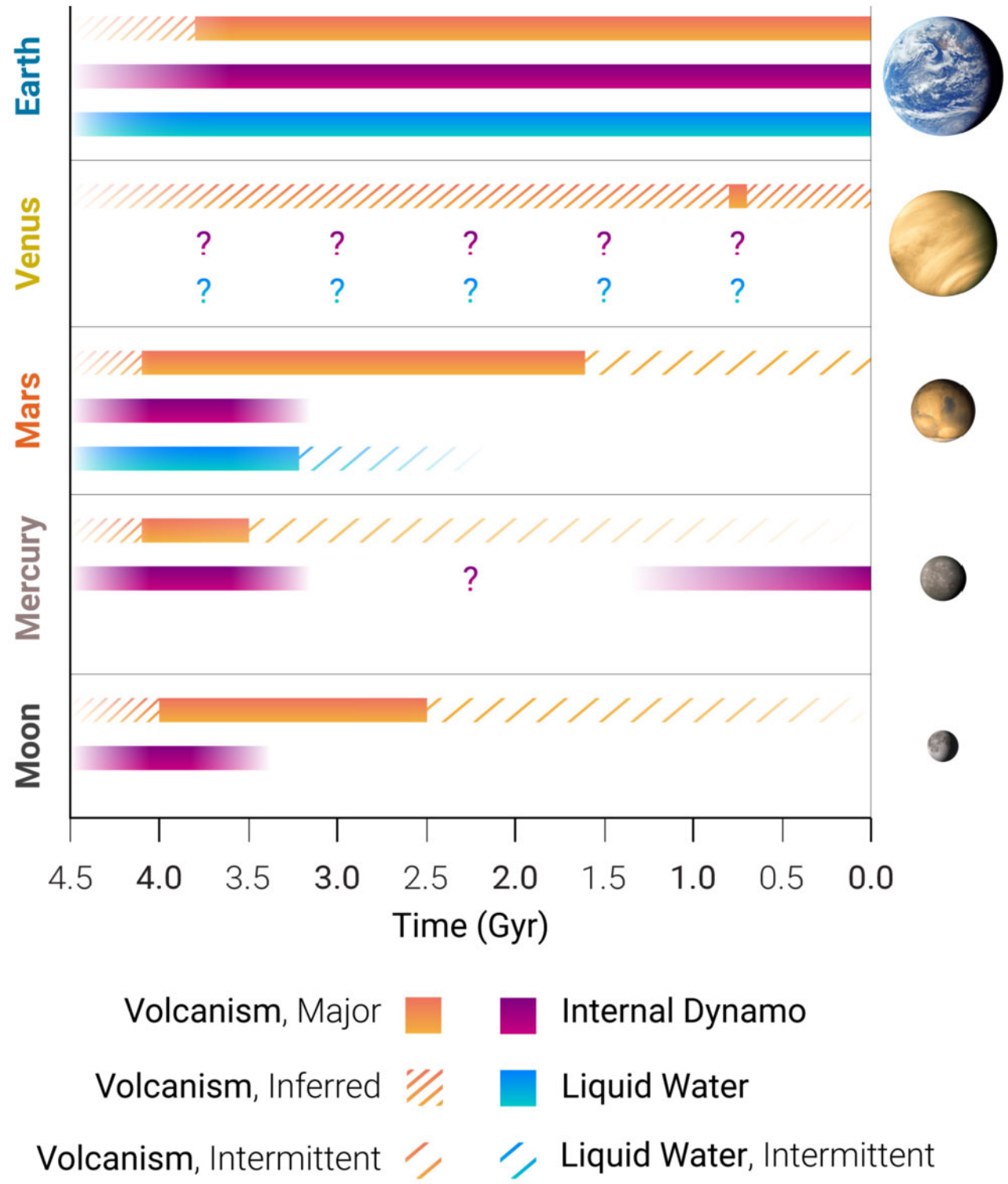

**Fig. 12** A schematic timeline of major volcanic activity, the generation of an internal magnetic field, and the presence of surface liquid water for the inner Solar System worlds. Major volcanism is defined here as continuous effusive plains volcanism; intermittent volcanism is taken to represent episodic activity; and inferred volcanism is the (chiefly effusive) activity that is thought to have occurred but is poorly preserved or absent from the geological record today (for Earth, Mars, Mercury, and the Moon), as well as activity that may be ongoing but has yet to be confirmed (for Venus). The major volcanism shown for Venus corresponds to a scenario under which the majority of the planet's plains were emplaced around 750 Ma. There is geological evidence for surface liquid water since almost the very beginning of Earth, and for approximately the first billion years of Mars' history. There may have been substantially smaller but nonetheless considerable volumes of liquid water on Mars for some time thereafter. Earth, Mercury, the Moon, and Mars had intrinsic dynamos in the ancient past; only Earth and Mercury are known to have internally generated fields today. Whether these modern magnetic fields are relatively geologically recent or have been continuous since early in Solar System history is unclear. The information available for Venus is, in general, extraordinarily lacking and so that world's geological history is essentially unknown to us. Adapted from Byrne (2020) and Cawood et al. (2025)

At home, the stark differences between Earth and Venus tell us that worlds within a single mass class can have widely divergent histories. Perhaps Venus' fate will one day befall Earth, either through catastrophic volcanic outgassing or under the influence of a steadily more luminous Sun (e.g., Wolf and Toon 2013). If so, then the contrast between these two worlds has less to do with stochasticity and more to do, again, with time.

And so we should bear in mind when we observe planetary systems of different ages, properties, and so on, that our own Solar System tells us that planetary evolution is not a readily predictable, linear avenue but a stochastic, wending trail. That we must be circumspect when interpreting rocky exoplanets. And that we should be prepared to be surprised.

Let that be the core lesson of this work. Only when sufficient time has passed will future planetary scientists be able to compare this writing with their state-of-the-art knowledge, and establish the extent to which it is prescient, and to which it is wanting.

**Acknowledgements** This study made use of NASA's Planetary Data System and Astrophysics Data System.

**Funding Information** P.K.B. acknowledges support from Washington University in St. Louis. C.M.G. acknowledges the support of the UK Science and Technology Facilities Council (grant no. ST/W000903/1) and the ETH Postdoctoral Fellowship. PAC was supported by Australian Research Council Grant FL160100168. J.C.D. was funded by FCT, IP/MCTES (PT) through national funds (PIDDAC) LA/P/0068/2020 and UID/50019/2025, and by the European Union's NextGenerationEU under projects UID/PRR/50019/2025 and UID/PRR2/50019/2025. F.M. acknowledges support from the Carnegie Institution for Science and the Alfred P. Sloan Foundation under grant G202114194.

## Declarations

**Competing Interests** The authors have no competing interests to declare that are relevant to the content of this article.

## Authors and Affiliations

**Paul K. Byrne[1,2] · Claire Marie Guimond[3,4] · Peter A. Cawood[5] · Michael J. Way[6,7,8] · Doris Breuer[9] · Tilman Spohn[9] · João C. Duarte[10] · Diogo L. Lourenço[4] · Francesca Miozzi[4,11,12] · Maëlis Arnould[13] · Nicolas Coltice[14] · Stephanie L. Olson[15]**

✉ Paul K. Byrne
paul.byrne@wustl.edu

1 Department of Earth, Environmental, and Planetary Sciences, Washington University in St. Louis, St. Louis, MO 63130, USA

2 McDonnell Center for the Space Sciences, Washington University in St. Louis, St. Louis, MO 63130, USA

3 Atmospheric, Oceanic, and Planetary Physics, Department of Physics, University of Oxford, Oxford OX1 3PU, UK

4 Department of Earth and Planetary Sciences, ETH Zürich, Zürich, Switzerland

5 School of Earth, Atmosphere and Environment, Monash University, Melbourne, Victoria 3800, Australia

6 Theoretical Astrophysics, Department of Physics and Astronomy, Uppsala University, Uppsala, SE-75120, Sweden

7 NASA Goddard Institute for Space Studies, 2880 Broadway, New York, NY 10025, USA

8 GSFC Sellers Exoplanet Environments Collaboration, NASA Goddard Space Flight Center, Greenbelt, MD 20771, USA

9 DLR Institute of Space Research, Berlin, Germany

10 Instituto Dom Luiz, Faculdade de Ciências, Universidade de Lisboa, Campo Grande, Lisbon, Portugal

11 Earth and Planets Laboratory, Carnegie Institution for Science, Washington, DC, USA

12 Department of Earth and Environmental Sciences, University of Pavia, Pavia, Italy

13 University Claude Bernard Lyon 1, ENSL, UJM, CNRS UMR 5276, LGL-TPE, Villeurbanne, France

14 UMR GéoAzur, Université Côte d'Azur, 250 Rue Albert Einstein, Valbonne, France

15 Department of Earth, Atmospheric, and Planetary Sciences, Purdue University, West Lafayette, IN 47907, USA